\documentclass[showpacs,preprintnumbers,amsmath,amssymb]{revtex4}
\usepackage{graphicx}
\usepackage{dcolumn}
\usepackage{bm}
\usepackage{epsf}

\newcommand{\beq}{\begin{equation}}
\newcommand{\eeq}{\end{equation}}

\newcommand{\eq}[1]{eq.(\ref{#1})}

\newcommand{\la}[1]{\label{#1}}
\newcommand{\ur}[1]{(\ref{#1})}

\begin{document}
\title{Kosterlitz-Thouless phase in systems of
one-dimensional strongly interacting fermions.}
\author{V. V. Afonin}
\affiliation{Solid State Physics Division, A.F.Ioffe  Institute,
194021 St.Petersburg, Russia}
\author{V. Yu. Petrov}
\affiliation{Theory Division, St.Petersburg Nuclear Physics
Institute, 188300, St.Petersburg, Russia}

\begin{abstract}
We present the ground state wave functions for
systems of one-dimensional interacting fermions. It is shown that
these systems undergo phase transitions similar to the
Kosterlitz-Thouless one independently of the interaction details.
In the limit of an  infinitely strong interaction the phase
transition turns into the usual  second  order phase transition in
a chiral  phase. The
temperature of the phase transition is calculated.
\end{abstract}

\maketitle

\section{Introduction}
\label{I}

One dimensional fermion systems were a subject of intensive
studies both in field theory and in condensed matter physics. As
was demonstrated by Tomonaga \cite{T} and Luttinger \cite{L} in
their pioneering papers the long range excitations of such a
system (under rather general conditions) can be expressed in
terms of non-interacting bosons. These degrees of freedom were
made explicit in the elegant method of bosonization proposed by
Mattis and Leeb \cite{ML}. The recent interest in this
field is mainly due to the development of submicron techniques
which allowed to produce very pure quantum wires. In such wires
only few levels (or sometimes even one) corresponding to the
quantization of electrons in perpendicular directions are
occupied. In spite of the fact that the number of the
experimental papers in the field is still not numerous (see, e.g.
\cite{B}, \cite{A}, \cite{Tak}, \cite{Deb})  the
systems under discussion are, in principle, accessible by
experiment.

The bosonization technique allows one to calculate all correlation
functions of the "density-density" type for  systems of
interacting fermions in one dimension. However it says nothing
about the  ground state of this system. The correlation functions
reveal a number of anomalies of the fermion system (see \cite{M},
\cite{E}, \cite{Shu}): they have oscillating contributions with
wave vectors equal to $2p_f$  or $4p_f$ which decay very slowly with
distance. In the literature these contributions were interpreted
as follows: the oscillations with Fermi momentum, $p_f$,
doubled were related to the Peierls instability (connected with
the charge density wave \cite{V}, \cite{E}) while the oscillations
with $4p_f$ frequency were interpreted as a marginal Wigner
crystal \cite{Shul}.
It is commonly believed that the system under discussion is a kind
of normal liquid, because quantum fluctuations destroy any order
parameter, and a phase with long-range order
is impossible even in the zero temperature region \cite{E}.
The common point of view  was put into the words: ``Luttinger liquid
is a normal (not symmetry-broken) metallic phase'' \cite{WM}
with gapless boson spectrum. This viewpoint
is usually supported by the
Landau theorem \cite{Lth} which states that in one dimension a long-range
order cannot exist \cite{theorem}. However two more
points should be taken into account:
\begin{itemize}
\item
It is well-known that the Landau theorem is applicable not
to all systems in low dimensions. Some of them appear to be in the
Kosterlitz-Thouless phase \cite{KT}. In such a phase the order
parameter density  tends to zero in an infinite system. However
a long-range order exists because  correlation functions decay as
some power of the distance. This means that the correlation  is
present in a whole specimen.  As a result the macroscopic
properties of the system are quite similar to  the system with
broken symmetry.  A Kosterlitz-Thouless phase can form if the
gapless excitations, which should be present in the system after
spontaneous symmetry breaking  due to the Goldstone theorem
(\cite{G}, \cite{G1}), {\em do not interact}.

At non-zero temperatures ($\Theta\neq 0$) this phase can appear in
two-dimensional systems; at $\Theta = 0$ a Kosterlitz-Thouless
phase is possible in one dimension. (The crossover temperature and
its relation  to the phase transition temperature
$\Theta_c$ will be considered below, see section \ref{r1}).
\item
The Goldstone theorem itself can be broken in one dimensional models
with Adler-Schwinger  anomaly. For that to happen the  interaction has to be
strong enough. As an illustration one can consider the massless
Schwinger model \cite{Sh}.  In the Coulomb gauge this model is a
particular case of the Luttinger model where the potential is a linear
function of the distance between electrons. This gives rise to the
Goldstone theorem violation: all excitations  have a gap in spite
of spontaneous breaking of the chiral symmetry. For this reason,
excitations cannot suppress the long-range order at
temperatures smaller than the gap. As a result one has a second
order phase transition  in one dimension even at  finite
temperature \cite{P}.
\end{itemize}
This makes the statement that the Luttinger liquid at $\Theta = 0$ is
in an unbroken phase doubtful.

In order to clarify this question
we calculate the wave functions of the ground states in the
Tomonaga-Luttinger model.
While the details of these wave
functions could depend on the interacting potential, qualitatively all
possible ground states reveal the same phenomenon.

In one dimension the Fermi surface reduces to two isolated points
in phase space ($p=\pm p_f$). One can neglect transitions between
these two points. This is a good approximation, at least if the
potential is a decreasing function of the  momentum transfer.
As a result, the number of electrons near every point ({\em left-}
and {\em right}-particles) should be conserved and the system acquires
complementary  ({\em chiral}) symmetry. It is
this symmetry which, as we will see, breaks down spontaneously
in the model.

The Fermi surface in one dimension is absolutely
unstable:  even infinitesimal electron-electron interaction leads to
instability of the electron ground state. The electron distribution
function is of the order of 1/2 near the Fermi
surface \cite{Lar}. This means that there is a hole located near
every electron.  Naturally, they attract each other and form a
kind of bound state consisting of a right electron and a left hole
($R\bar L$-pair). This is quite similar to the formation of a Cooper
pair in a superconductor but the quantum numbers of the bound state
are different: instead of non-zero electric charge it has non-zero
chirality \cite{par}.

Of course, this fact itself is not enough to speak about a new
correlated phase. We will  check  using the explicit ground state
wave function constructed in the paper that long-range order is
indeed present in the system. At low temperatures
Luttinger liquid undergoes a phase transition similar
to the Kosterlitz-Thouless one.

We will see that in the  limit of the infinitely strong interaction
and at zero temperature the system is in a phase with broken chiral
symmetry and non-vanishing order parameter. The properties of the
Luttinger liquid in this limit are quite analogous to the properties
of the massless Schwinger model where the spontaneous breakdown of the
chiral symmetry is well-known. This can be expected as
the interaction in the Schwinger model is also infinite (it grows
with distance). On the other hand, in contrast to the Schwinger
model, the spectrum of the Luttinger model remains gapless. The
only effect of the interaction on the spectrum is the renormalization
of the Fermi velocity $v_f^c$ which is proportional to the strength of
the interaction.

If the interaction in the Luttinger model is considered to be finite, the
order parameter vanishes (in the infinite system). The system appears
to be
in a Kosterlitz-Thouless phase with correlators which decay as
some power of the distance. However the properties of this phase
are rather close to those for a system with non-zero order parameter.
The Kosterlitz-Thouless system transforms smoothly  to the phase with
non-zero order parameter as the coupling constant increases.

Another matter is the temperature $\Theta_c$ of the phase transition
to the unbroken phase. This temperature is always of the order of
the gap in the spectrum. For this reason $\Theta_c$ is finite for
the Schwinger model and of order of $v_f^c/L$ for the Luttinger
liquid (where $L$ is the size of the system). In other words $\Theta_c$
tends to zero for an infinite system. However  for real
systems $\Theta_c$ is not so small. Indeed, if
one takes $v_f \sim 10^7 cm/sec$ and $L\sim 10^{-4} cm$
then $\Theta_c \sim 1^{\circ}$ K $\cdot v_f^c/v_f $. This means
that the theory based on the limit $L\to\infty$ can be inadequate.
Of course, in order to discuss $\Theta_c$ for
finite size systems we have to define the notion of phase transition
in this case. This is done in Section \ref{r1}.

We will see that a non-zero order parameter is the reason of the
anomalies in the correlation functions mentioned above. Anomalies
exist even at temperatures higher than $\Theta_c$ owing to
fluctuations of the chiral phase in the phase with non-broken
chiral symmetry. There is a clear-cut distinction between
the chiral phase and the Peierls's phase. The latter is the second order
phase transition in the phonon system and the chiral symmetry of
electrons is broken explicitly, in the Hamiltonian. On the
contrary the chiral transition originates from the spontaneous
symmetry breaking in the electron system.

Finally, let us discuss  effects which may indicate the  existence
of a condensate.  Obviously a charge-neutral condensate can not
reveal itself in experiments associated with charge transfer.
However it contributes to the effects concerned with energy
currents and should not transfer heat. Hence one can think about
thermal anomalies related to the condensate. In fact, we keep in
mind the effect similar to the thermomechanical effect in
superfluid helium. (The temperature decreases with increase of the
superfluid mass \cite{L6}). We plan to discuss this problem in a
separate publication.

The paper is organized as follows. In Section \ref{G} we introduce
the Hamiltonian, definitions of  left- and right-particles and so
on. We present our results and discussion in  Section \ref{r1}. We
relegate the derivation to the next sections as the calculation is
a bit cumbersome.

\section{Notations and general equations.}
\label{G}

We begin with the usual expression for the Hamiltonian of interacting spinless
electrons in one dimension:
\begin{eqnarray}
H = \int dx  \hat \Psi^{\dag}\left( x \right)\left(-{1\over 2m}
\partial ^2_x  - \mu \right) \hat \Psi \left( x  \right)\nonumber\\   +
\int dx dy   \hat \Psi^{\dag}\left( x \right) \hat \Psi \left( x  \right)
V\left( x-y \right) \hat \Psi^{\dag}\left( y \right) \hat
\Psi \left( y  \right) .
\label{g1}
\end{eqnarray}
Here $\hat \Psi \left( x  \right)$ is the electron field, $m$ is the electron
mass and $\mu$ is the chemical potential. $V\left( x-y \right)$
is the electron-electron interaction which we will discuss  below ($\hbar = 1$).

As usual  let us separate in the electron wave functions left- and right-
particles \cite{TS}:
\beq
\hat \Psi \left( x\right)=\exp\left( ip_f x\right)\hat
\Psi _R\left( x \right) +\exp\left( -ip_f x\right)\hat \Psi_L \left( x
\right) .
\label{g2}
\eeq
It is implied here that the
wave functions  $\hat \Psi _{R,L}\left( x\right)$
are varying over distances much larger than $1/p_f$. We also restrict
ourselves to the Tomonaga-Luttinger model and assume that the characteristic
scale of the potential is large as compared to $1/p_f$. Lastly, for
simplicity we consider only an electrically neutral system where the a positive
charge of ions is distributed homogeneously along the channel.

Let us proceed to the electron-hole
representation for the right (left) particles:
\beq
\hat \Psi_{R,L} \left( x  \right)= \int \limits_0^{\infty} {dp\over2 \pi}
\left( \exp\left( \pm ipx \right)\hat a_{R,L}\left( p  \right) +
\exp\left( \mp ipx \right)\hat b_{R,L}^{\dag}\left( p  \right)\right)=\\
\hat a_{R,L}\left( x  \right) +
\hat b_{R,L}^{\dag}\left(x  \right) .
\label{abr}
\eeq
Here $\hat a^\dag (\hat a)$ and $\hat b^\dag (\hat b)$  are creation
(annihilation) operators for electrons and holes.
The Hamiltonian of \eq{g1} can be written for neutral system in terms
of electron density operator
$$\varrho\left(
x  \right)=\varrho_R\left( x  \right)+ \varrho_L\left( x  \right)
$$
in the form
\beq
H = \int dx \left[ \hat \Psi^{\dag}_R \left( x
\right) v_f\left( -i \partial_x \right) \hat \Psi_R \left( x  \right) +
 \hat \Psi^{\dag}_L \left( x \right) v_f i\partial_x
\hat \Psi_L \left( x  \right) \right]\\   +
\int dx dy \varrho \left( x  \right) V\left( x-y \right)
\varrho\left( y  \right).
\label{g5}
\eeq
Here $v_f$ is the Fermi velocity, and
\beq
\varrho_{R,L}\left( x  \right)=
\hat a_{R,L}^{\dag}\left( x  \right)\hat a_{R,L}\left( x  \right) -
\hat b_{R,L}^{\dag}\left( x  \right)
\hat b_{R,L}\left( x  \right)+ \hat a_{R,L}^{\dag}
\left( x  \right)\hat b_{R,L}^{\dag}\left( x  \right)
+ \hat b_{R,L}\left( x  \right)\hat a_{R,L}\left( x  \right) .
\label{g4}
\eeq
The expression \ref{g5} coincides with the
Hamiltonian of the Tomonaga-Luttinger model (see, e.g., Ref.~\cite{TS}).

Let us discuss now the electron-electron interaction $V(x-y)$. Its
form depends on the relation between the  usual 3D  screening
radius $R_D$  (for simplicity we will consider the case of
 Debye screening) and the transverse size of
the channel $d$. Indeed one has to take into account that the
electrons are one-dimensional only for distances
$|x-y|$ which are much larger than $d$. Therefore if $R_D\ll d$
one can use a point-like interaction $ V_0\delta \left( x-y \right)
$\cite{Back}. This is the case for metals \cite{Bohr}. In the opposite
case $R_D\gg d$, semiconductor, one should use an ordinary Coulomb
potential. In one dimension the Fourier transform of the Coulomb
potential is logarithmically divergent. This divergence is
regularized by the smaller of two quantities, either $p_f$ or the inverse
channel size, $1/d$. Thus  \cite{cate}
\begin{itemize}
\item At $R_D\ll d$
\begin{equation}
V\left( p \right)= V_0 ,\\
\label{g7a}
\end{equation}
\item
At $R_D\gg d $
\begin{equation}
V\left( p \right)= \left\{
\begin{array}{cc}
2e^2\log\left({2p_f\over | p |}\right)& \mbox{\rm for}\;  1/ p_f\gg d\\
2e^2\log\left({1\over | p |d}\right) & \mbox{\rm for} \; 1/p_f\ll d\\
\end{array}
\right.
\label{g7}
\end{equation}
\end{itemize}

The Hamiltonian of the Luttinger model, \eq{g5}, presented in terms of
electrons and holes is completely defined without any additional
regularization of electron operators. In particular,
the commutator of the $R$- and  $L$-densities in this representation
reproduces the well-known Schwinger anomaly \cite{Sh}:
\beq \left[
\varrho_{R,L}\left( x  \right),\varrho_{R,L}\left( y
\right)\right]= \pm {i\over2\pi}{\partial \over \partial x}
\delta\left( x-y \right) .
\label{g8}
\eeq
These relations are the
starting point of the bosonization technique. Usually one derives
\eq{g8} regularizing the product of $\Psi$-operators by a small
shift of their arguments \cite{Kog}. This is not necessary, however,
in the electron-hole representation \cite{P} as the
creation-annihilation operators for R,L electrons
($\hat a^\pm_{R,L}(x)$) and holes ($\hat b^\pm_{R,L}(x)$) are non-local in
the coordinate space:
\begin{equation}
\{ \hat a_R^{\dag}\left( x  \right) ,\hat a_R\left( x_1  \right)
\}    = \{\hat b_R^{\dag}\left( x  \right) ,\hat b_R\left(
x_ 1  \right)\} =  {1\over2\pi i}\cdot {1\over x-x_1-i\delta}
\label{g9}
\end{equation}
\begin{equation}
\{
\hat a_L^{\dag}\left( x  \right) ,\hat a_L\left( x_1  \right) \}  =
\{ \hat b_L^{\dag}\left( x  \right) ,\hat b_L\left( x_1
\right)\} =  {1\over2\pi i}\cdot {1\over x_1-x-i\delta} .
\label{g10}
\end{equation}
(in momentum space these anticommutators are $\delta$-functions).
Using these anticommutators for the densities of right and left
electrons $\varrho_{R,L}\left( x\right)$ in the form of \eq{g4} we
immediately reproduce the Schwinger anomaly.
This means that, being formulated in the electron-hole
representation, our theory is completely determined without any further
redefinition of the density operators.

The Hamiltonian  \ref{g5} is invariant both under {\em vector} transformations:
\beq
\Psi_R(x)=e^{i\alpha_V}\Psi_R, \quad
\Psi_L(x)=e^{i\alpha_V}\Psi_L
\eeq
and under {\em chiral} transformations
\beq
\Psi_R(x)=e^{i\alpha_c}\Psi_R, \quad
\Psi_L(x)=e^{-i\alpha_c}\Psi_L .
\label{chiral}
\eeq
The first invariance leads to conservation of  electrical charge
(the number of left electrons plus the number of right ones), the
second means that chiral charge (the difference of numbers of right and
left electrons) is conserved. However we will see below that the
ground state of the model is constructed in such a way that the second
symmetry can be spontaneously broken.

\section{Approach, Results and discussions}
\label{r1}

The standard approach to systems of many particles is based on  Green
functions. The one-particle Green function gives the information about the
spectrum of excitations; the many-particle Green functions
allow one to calculate different correlation  and response functions. Of
course, the Green functions give some information about the wave functions
of the states but this information is indirect.

In principle, one can obtain the wave functions of  stationary
states (and, in particular, of the ground state) by solving the
corresponding Schr\"odinger equation directly. However for
systems with infinite number of degrees of freedom this equation is
too complicated. A more practical approach can be based on the
evolution operator \cite{Feynman-Hibbs} which is a sum~:
\begin{equation} S\left( T \right) =
\sum_{m,n} |n> <n|\exp{-iHT}|m> <m| .
\label{s1}
\end{equation}
Here $|n>$ are the exact wave functions of the Hamiltonian $H$ in
the second quantized representation, $T$ is the time of
observation. The evolution operator determines  the evolution of an
arbitrary initial wave function ($<m|$) from the time $t=0$ up to
final states  $|n>$ (at $t=T$). (We imply from now that the
Schr\"odinger representation for operators with time-dependent
wave-functions is used.)

Formula (\ref{s1}) suggests the general method to obtain wave
functions. One has to calculate first the evolution operator and
represent it as a sum of time-dependent exponents. The coefficients
in front of these exponentials are products of exact wave functions
and their complex conjugates. In order to extract the ground state
wave function one has to take the limit $T\to\infty$ (we add
an infinitesimal imaginary part to the energy). Proceeding to
Euclidean time ($T\rightarrow -i/\Theta$ we see that evolution
operator determines the density matrix for the equilibrium system
at non-zero temperature (see end of the Section \ref{vo}).

The advantage of this method is that the evolution operator can be
written explicitly as a functional integral with definite boundary
conditions (see \eq{s2} ). This functional integral is rather
simple for the Luttinger model with Hamiltonian (\ref{g5}) and can
be calculated exactly. This allows one to construct wave functions
of all states in the model and, in particular, the ground state.  This
will be quite enough in order to demonstrate the symmetry breaking
and to calculate the temperature of the phase transition.

We will keep the size of the system finite. This is important
not only to regularize infrared divergences in the system but mostly
because of the peculiar situation with the temperature $\Theta_c$
mentioned in the Introduction. However, first, we
have to define the concept of phase transition in finite size systems.

Usually, the critical temperature is defined as a point where thermodynamic
quantities have a singularity. Of course, this is the case only in
a infinite system because all singularities smear out if the size of the
system is finite. The same is true for the coherence length --- it
cannot be larger than the size of the system.

In this paper we will adopt the point of view suggested by Landau
in order to describe the second order phase transitions \cite{LY}.
He introduced  the order parameter as the main quantity for the
description of phase transitions related to the spontaneous
symmetry breakdown. By definition, the  order parameter is zero in
the high symmetry phase (with the same symmetry as the Hamiltonian)
and non-zero in the phase with broken symmetry. In the case of
chiral symmetry \eq{chiral} the following quantity can serve as an
order parameter:
\begin{equation} \Delta = \int dx
<\Omega |\hat a^{\dag }_R\left( x + \delta x/2 \right) \hat
b^{\dag}_L\left( x -\delta x/2 \right) |\Omega >|_ {\delta x \to 0} .
\label{del1}
\end{equation}
This quantity is not invariant under transformations of
\eq{chiral} and should be zero if the chiral symmetry remains
unbroken. Note that we use a macroscopic order parameter (integral
over the whole size of the system). In the broken phase this
quantity is proportional to the volume of the system.

The Kosterlitz-Thouless phase represents the intermediate case when
\beq \Delta \sim L^{\alpha_T}, \quad 0<\alpha_T < 1, \la{KTcase}
\eeq so $\Delta$ is still infinite in the thermodynamic limit
while the {\em density} or order parameter $\Delta/L$ vanishes.
Let us point out that $\Delta$ appears to be non-zero even at
$\Theta>\Theta_c$ due to fluctuations of the broken phase in the phase
with higher symmetry. What is important  for the latter case is the
fact that $\Delta$ does not increase with $L$.

Intensive thermodynamic quantities
remain smooth for finite size system even at the
point of the phase transition. However what matters is the fact that
they depend explicitly on the system size and tend to infinity
(or acquire a jump) at  $L\rightarrow\infty$.

Usually one proves that the system is in the Kosterlitz-Thouless
phase by investigating the behaviour of the four-fermion correlator
which does not break the chiral invariance ( below in section
\ref{K}we will consider such a correlator, namely the probability
to find an $R\bar{L}$-pair at
a large distance $r$ from the $L\bar{R}$ pair).  If such a
correlator decreases sufficiently slowly with the
distance, the system is in the Kosterlitz-Thouless phase. The
limiting case when the correlator remains constant at large
distances corresponds to a non-zero density of the order parameter and
ordinary broken symmetry. In fact, this definition of the
Kosterlitz-Thouless phase is equivalent to our definition given
above  but definition (\ref{del1}, \ref{KTcase}) is more convenient  for us.

In one dimension the Kosterlitz-Thouless phase can exist only at
$\Theta=0$ or, to be more precise, for the temperatures which tends
to zero at $L\rightarrow\infty$. There is no need in microscopic
theory in order to estimate $\Theta_c$:  one can use general
phenomenology applicable to all Kosterlitz-Thouless systems (see, e.g.
\cite{TS}).

Let us assume that the chiral symmetry is indeed spontaneously
broken in the Luttinger model (of course, this can be proved only
in microscopic theory).  According to the Goldstone theorem the chiral
phase $\alpha_c$ (the phase of the $\Psi^+_R\Psi_L$ operator) becomes
a massless boson field. In the long-range limit only fluctuations
of this field are relevant and its effective Lagrangian reduces
to (Euclidean time $\tau=it$ is used, as we want to consider
non-zero temperatures below):
\beq
S_{eff}[\alpha_c]=\frac{V^2}{2}\int d\tau dx
\left[(\partial_t\alpha_c)^2+ (w\partial_x\alpha_c)^2 \right] ,
\eeq
where $V$ and $w$ are  phenomenological constants (calculable
in microscopic theory).

To judge if the system is in the Kosterlitz-Thouless phase it is
enough to consider the behaviour at large distances $|x-y|$ of the
chirality conserving correlator:
\beq
F(x-y)=<\Psi^+_R(x)\Psi_L(x),\Psi^+_L(y)\Psi_R(y)> .
\label{correlator1}
\eeq
One can neglect fluctuations of the
modulus of the operator $\Psi^+_R(x)\Psi_L(x)$ (as well as higher
derivatives in the effective action for chiral phase).
 Then the
correlator (\ref{correlator1}) reduces to:
\beq F(x-y)=const \int
{\cal D}\alpha_c \exp(-S[\alpha_c]) e^{2i\alpha_c(x)} e^{-2i\alpha_c(y)}
\eeq
Calculating the latter integral at $\Theta=0$  we get:
\beq
F(x-y)\sim \exp{\left[ 2V^{-2}\int {dk_0 dk\over \left( 2\pi
\right)^{2}} {\sin^2 \left[ 1/2\left( {\bf k, x-y}
\right)\right]\over \left( k_0^2 +w^2{\bf k}^2 \right)}\right] } .
\label{cd}
\eeq
The two-dimensional integral ( one space one time
dimension ) in the exponent (\ref{cd}) diverges logarithmically and hence
\beq F(x-y) \sim { V^2 \over \left( k_{max} |x-y| \right)^{1/2\pi
V^2w}} .
\label{thetanol}
\eeq
This proves the existence of
Kosterlitz-Thouless phase at $\Theta=0$.

If the temperature is non-zero the integral over $k_0$ in
\eq{cd} should be replaced by a sum over discrete
values $k_0=2\pi n \Theta$ ($n$ integer). At high temperatures
only the term with $n=0$ survives at large distances and we are
left with an one-dimensional integral in $k$ which leads to the
correlator exponentially decreasing with  distance:
\beq
F(x-y)\sim V^2 \exp{ \left( -{\Theta\over 2\pi V^2w^2} |x-y|
\right)}
\label{cd1a}
\eeq
Clearly this correlator describes the
unbroken phase.

A power - like behavior of the correlator of (\ref{thetanol}) is valid
in the region $$|x-y|< w/\Theta.$$ For
$$ \Theta < \Theta_c\equiv
w/L $$
this takes place for the whole specimen, i.e. the system is
in broken a phase. The temperature $ \Theta_c $ is a temperature of
the phase transition.

 One can recognize in this estimate the excitation energy
with the smallest momentum possible in a finite size system. In the
Luttinger model this energy is equal to $ \omega_{min} = 2 \pi
v_f^c/ L$ with renormalized Fermi velocity $ v_f^c = v_f \sqrt {1
+V_0/ \pi v_f}$ \cite{E}. We see that if the spectrum of
excitations is gapless (like in the Luttinger model), then the
temperature of phase transition is inversely proportional to the
specimen length. This result can be obtained in the microscopic
theory as well (see Section \ref{vo}).

Turning to the microscopical theory we begin with the simplest case:
the short range potential in the limit of an infinitely strong interaction
(\ref{g7a})
\beq
 \frac{\pi v_f}{V_0}  \ll 1 .
\label{rv4}
\eeq
In the leading order in this parameter the evolution operator
appears to be very simple and the wave function of the
ground state can be represented in close form. In the temperature
region
\begin{eqnarray}
\Theta_{chiral}= {2\pi v_f\over L}\ll \Theta\ll \Theta_c=  \omega
_{min},
\label{rt1}
\end{eqnarray}
the ground state wave function is of the
form \cite{matrix}:
\begin{eqnarray} |\Omega >_{\theta} = \sqrt
{Z_0}\exp \left[\int dx \exp\left( i\theta  \right)\hat a^{\dag
}_R \left( x  \right) \hat b^{\dag }_L\left( x  \right) + \int dy
\exp\left( -i\theta  \right) \hat a^{\dag }_L\left( y  \right)
\hat b^{\dag }_R \left( y  \right) \right] | F > .
\label{rt5}
\end{eqnarray}
Here $|F>$ is the filled Fermi sphere and $Z_0$ is the normalization
coefficient. There is an infinite set of degenerate ground states
labeled by the continuous parameter $\theta$ which has the meaning of
the order parameter phase.

One can see  the symmetry breaking immediately because the
wave function (\ref{rt5}) is not invariant under the chiral transformation
(\ref{chiral}). Besides one can check directly that
$\Delta \propto L$  \cite{rem7}. It means that there is  a
second-order phase transition in this limit.

The wave function of (\ref{rt5}) is a mixture of states with different
chirality. (We assign chirality $+1$ to a right electron and a left
hole and $-1$ to their counterparts. So bosons
in \eq{rt5} are neutral in terms of electric charge but have a
{\em non-zero chirality}, $\pm 2$ ). Such a ground state implies
that the states with different chirality are all degenerate in
energy. This is a typical situation for  systems with condensate:
the addition of one pair to the condensate does not cost any energy.
However this degeneracy is possible only if the size of the system
is large enough, namely, we will see that it should be $L\gg
L_{min} \sim 2\pi v_f/ \Theta$. At $L\le L_{min}$ the ground state
has fixed chirality (equal to zero) (see \eq{v7}) and the order
parameter $\Delta$ vanishes, i.e. there is no spontaneous symmetry
breaking. These considerations put a lower bound on
the temperature region
where a chiral phase can exist: $\Theta\gg
\Theta_{chiral}$.  So, $\Theta_{chiral}$ is the degeneration
temperature.

Let us estimate the density of chiral pairs in the ground state.
The wave function (\ref{rt5}) implies that {\em all} electrons are bound
to the pairs. Hence
the density of $R\bar{L}$ coincides with the density of
$R$-electrons:
\beq N_R \left( p \right) = _{\theta}<\Omega |\hat
a^{\dag }_R\left( p \right) \hat a_R\left( p \right) |\Omega
>_{\theta} = L/2.
\la{pairs}
\eeq (see \cite{rem7}).
This quantity
reflects a well-known fact: the distribution function of
electrons  is of the order of 1/2 near the Fermi surface
\cite{Lar}.  If the interaction is infinitely large all electrons
and all holes are bound in exciton-like  pairs. As a result we
get the value (\ref{pairs}) which is maximal possible.

In the model under consideration  $N_R  \left( p \right)$ is
momentum independent and the total number of pairs $N_R$ diverges
at large $p$.  (This is the defect of the point-like
electron-electron interaction.)  The sum over all states
should be restricted either by $p_f$ or, at $p_fd\gg1$, by the
inverse size of the channel because at larger $p$ electrons cannot
be considered as one - dimensional (see \cite{Bohr}). As a result
$$
N_R  \sim {L\over 4\pi d} .
$$
 Thus the number of pairs $N_R$ is only a small
fraction of the total number of electrons ($ Lp_f/2\pi $). This does
not mean, of course, that in this case the Luttinger liquid  behaves
like a normal one. The reaction of the system to the slowly varying
external fields is determined completely by the electrons near the
Fermi surface which are all paired. This  situation reminds
superfluid helium where (even at zero temperature) the density of
the condensate is only few percents of the total one. Nevertheless the
whole mass of helium is superfluid \cite{L1X}.

Let us proceed with the region of high temperatures: $\Theta\gg
\Theta_c=\omega _{min}$.  In this region the macroscopic order
parameter (\ref{del1}) is proportional not to the volume of the
system but to the some characteristic length
$\zeta\left(\Theta\right)=v_f ^c/\left( \Theta - \Theta_c \right)
$ (see the end of section \ref{vo}) and the density of the order
parameter $\Delta/L$ vanishes in the limit $L\to\infty$ as it
should be. Hence the temperature $\Theta_c$ has indeed the meaning of
the temperature of phase transition from the symmetrical phase
to the phase with broken chiral symmetry.

The length $\zeta  \left( \Theta \right)$ plays the role of {\em
coherency length} in our system. At lengths less than
$\zeta  \left( \Theta \right)$
the wave function of the system
coincides with the coherent exponent (\ref{rt5}). However at larger
distances the order disappears.

The macroscopic order parameter $\Delta$ can be non-zero even
in the symmetrical phase due to fluctuations of the broken
to the unbroken phase. What matters is the behaviour of
$\Delta$ with the size of the system.
If $ \Delta $ does not increase with $L$
($\Delta\propto\zeta$ with $\zeta$ finite),
we deal with unbroken phase, where
$\Delta$ increases with $L$
the long range order
appears (and $\zeta\sim\L$). Such a dependence $\zeta$ on $L$  can be
considered as the {\em definition} of  symmetry breakdown for a
finite size system too.

On the other hand it is obvious from such a definition
that a temperature of the phase transition in the finite
size systems can be defined only up to  $1/L$ corrections and
the phase transition is smooth within the $1/L$ region around the
temperature of the phase transition. In the
Luttinger model, where the temperature $\Theta_c$ itself is of $1/L$
order, we can define $\Theta_c$ only up to a factor of order
of unity. This is the price we have to pay for considering a
phase transition of a large but finite size system. However
there is still a clear-cut distinction between the case with a
correlation length of the order of the size of the system (broken phase)
and the case when $\zeta\ll L$ (unbroken phase).

As was already pointed out in the Introduction, the case
of the infinitely strong interaction is very special.
We will see that if the interaction is finite, then the macroscopic
order parameter $\Delta$ grows with the system size
but more slowly than $L$ (at $\Theta\ll \Theta_c$). In the case of a short
range potential (see Section \ref{K}) $\Delta$ behaves as some power
of $L$. This  corresponds literally to the definition of the
Kosterlitz-Thouless phase. If we consider the potential of Coulomb
type, then $\Delta$ depends on $L$ in a more complicated way (section
\ref{cul}) but still $\Delta$ increases with the size $L$. Physically,
this case is quite similar to the usual Kosterlitz-Thouless one.

To summarize, the Luttinger model at $\Theta<\Theta_c$ is always  in
Kosterlitz-Thouless phase with broken chiral symmetry. At
$\Theta \sim \Theta_c$ it undergoes
a phase transition which in the limit of infinite interaction
turns into the ordinary second order phase transition.

\section{Ground state of the Tomonaga-Luttinger model.}
\label{cg}

The evolution operator (\ref{s1}) of the quantum system can be represented
as a functional integral with definite boundary conditions (see,
e.g. \cite{Feynman-Hibbs}). Usually one derives this
representation for boson systems, for the sake of completeness we
give in Appendix \ref{ev} the derivation for fermions.

The theory with arbitrary electron-electron interaction can be
reduced to a theory in an external field by means of the
Hubbard-Stratonovich transformation \cite{Hub}(see below, \eq{s10}).
 One has to integrate over the value of the
external field in order to return to the original 4-fermion
interaction. For this reason we consider first the evolution
operator for one-dimensional electrons placed into an external
field $\Phi (x,t)$. It is of the following form:
\begin{equation}
\hat S \left( \Phi \right)=\int_{\left( \overline{\Psi} ,\Psi\right) }
{\cal D}\Psi {\cal D} \overline{\Psi}
\exp{{\cal S}
\left( \overline{\Psi} ,\Psi\right)}.
\label{s2}
\end{equation}
Here $\overline{\Psi} ,\Psi$ is the electron field (Grassmann variables)
and ${\cal S}$ is the action:
\begin{eqnarray}
{\cal S} = i \int_0^T dt \int dx \overline{\Psi}_R\left( x,t \right)
\left[ i\partial_t -v_f i\partial _x  +\Phi\left( x,t\right)
\right] \Psi_R \left( x,t\right)\nonumber\\
 + \left( R,v_f \leftrightarrow  L,-v_f,\right) .
\label{s3}
\end{eqnarray}

Integration over $\overline{\Psi} ,\Psi$ in \eq{s2} is performed
with given boundary conditions at $t=0$ and $t=T$:

At $t  \to +0 $
\[
\Psi_{R,L} \left( x,t \right) = \hat a_{R,L}\left( x  \right) +
\mbox {arbitrary negative frequency part}
\]
\[
\overline{\Psi}_{R,L}\left( x,t \right) = \hat
b_{R,L}\left( x  \right)+
\mbox {arbitrary negative frequency part}
\]

At  $t  \to T-0$
\[
\Psi_{R,L} \left( x,t \right) = \hat b^{\dag }_{R,L}\left( x  \right) +
\mbox{arbitrary positive frequency part}
\]\beq
\overline{\Psi}_{R,L}\left( x,t \right) = \hat a^{\dag }_{R,L}\left( x
\right)+\mbox {arbitrary positive frequency part}
\label{s4}
\eeq

The creation operators of electrons and holes $\hat a^+, \hat b^+$ are the
variables which enter the wave functions of the states in the sum
of \eq{s1}. The annihilation operators $\hat a,\hat b$ enter the conjugate wave
functions. They anticommute: $\{\hat a,\hat a^+\}=\{\hat b,\hat b^+\}=0$ since they
belong to
different instances of time  as long as
one calculates the evolution operator (see Appendix \ref{ev} for
details).

It is  possible to separate  explicitly the dependence on
creation-annihilation operators for the evolution operator in a
given external field determined by the functional integral  (\ref{s2}).
Let us introduce new  integration variables:
\begin{eqnarray}
\Psi _{R,L} = \Psi^0_{R,L} + \chi_{R,L}\nonumber\\
\overline{\Psi}_{R,L} = \overline{\Psi}_{R,L}^0 + \overline{\chi}_{R,L}.
\label{s4a}
\end{eqnarray}
The saddle-point fields $\Psi^0_{R,L}$ are supposed to obey the Schr\"odinger
equation in the external field $\Phi(x,t)$ with given boundary
conditions (\ref{s4}).
The ``quantum" fields $\chi_{R,L}(x,t)$ are arbitrary but obey zero  boundary
conditions:
$\chi_{R,L}\left( 0 \right) =\chi_{R,L}\left( T \right) = 0 $.

The solutions $\Psi^0_{R,L}$ can be represented in terms of the Feynman Green
function in the finite time $G_{R,L}$ which is defined as follows.
It is a solution of the Schr\"odinger equation:
\begin{equation}
\left[ i\partial_t \mp v_f i\partial _x  +\Phi\left( x,t\right) \right]
G_{R,L}\left( x,t;x_1,t_1\right) =
i\delta^{(2)}\left( x-x_1,t-t_1\right)
\label{s5}
\end{equation}
with the following boundary conditions: at $t \to +0$ the Green function
$G_R(x,t,x_1,t_1)$ should coincide with the Green function
of  free fermions in the lower semiplane of the complex variable x (being
arbitrary in the upper semiplane). At $t \to \left( T-0\right)$  it coincides
with the free Green function in the upper semiplane. For the Green function
of left electrons $G_L(x,t,x_1,t_1)$ one has to exchange upper and
lower semiplanes.

The free Feynman Green function is equal to~\cite{M}:
\begin{equation}
G_{R,L}^0 \left( x,t;x_1,t_1\right) = {1 \over 2\pi i} \left[ v_f
\left( t-t_1\right) \mp \left( x-x_1\right)
-i\delta {\rm sign}\left( t-t_1\right) \right] ^{-1}.
\label{s6}
\end{equation}

In one dimension the Schr\"odinger equation (\ref{s5}) can be solved
for an arbitrary external field $\Phi(x,t)$:
\[
G_{R,L}\left( x,t;x_1,t_1\right)=G_{R,L}^0\left( x,t;x_1,t_1\right)
\times
\]\beq
\times\exp \left[ i \int _0^T dt' \int dy  \Phi (y,t')\left(
G_{R,L}^0\left( x,t;y,t'\right) - G_{R,L}^0\left( x_1,t_1;y,t'\right)\right)\right] .
\label{s6a}
\eeq

Now it is easy to verify that the saddle point fields $\Psi^0_{R,L}$
can be expressed in terms  of these Green functions as follows:
\begin{eqnarray}
\Psi^0_{R,L}\left( x,t \right) = \int dx'  \left[ G_{R,L} \left( x,t;x',0
\right) \hat a_{R,L}
\left( x ' \right)   -
G_{R,L} \left( x,t;x'T \right) \hat b^{\dag }_{R,L}\left( x ' \right)\right]
\nonumber\\
\overline{\Psi^0}_{R,L}\left( x,t\right) = -\int dx' \left[ G_{R,L}
\left( x',0;x,t\right)\hat b_{R,L}
\left( x ' \right)   -
G_{R,L} \left( x',T;x,t\right) \hat a^{\dag }_{R,L}\left( x ' \right)
\right] .
\label{s7}
\end{eqnarray}
In order to check that these fields obey the required boundary conditions
let us note that $\hat a_R\left( x  \right)$ and $\hat b_R\left( x  \right)$
are regular in the upper semiplane (see \eq{abr}). Therefore the
positive frequency part at $t\to +0$ of $G_R(x,t,x_1,t_1)$ is
determined by the pole contribution at $x'= x + i \delta $ and is equal
to  $\hat a_R\left( x  \right)$ as it should be. The second term in
\eq{s7} gives the negative frequency part which is arbitrary. Similarly,
one can check the boundary condition also at $t\to \left( T-0\right) $.
Inside the time interval $(0,T)$ the saddle point fields satisfy the
Schr\"odinger equation as is seen from \eq{s5} for the Green
functions.

The contribution of the saddle-point field to the action is:
\begin{eqnarray}
{\cal S}_0 = \sum_{i=R,L} \int dx dx'  \left[
\hat b_i \left( x ' \right) G_i\left( x',0;x,\epsilon \right)
\hat a_i\left( x  \right) +
\hat a^{\dag }_i\left( x ' \right) G_i\left( x',T;x,T-\epsilon\right)
\hat b^{\dag }_i
\left( x  \right) \right.\nonumber\\
-\left. \hat a_i^{\dag}\left( x ' \right) G_i\left( x',T;x,0\right)
\hat a_i\left( x  \right) -
\hat b_i\left( x ' \right) G_i\left( x',0;x,T\right) \hat b_i^{\dag }
\left( x  \right) \right] .
\label{s8}
\end{eqnarray}
Since the saddle point fields obey the Schr\"odinger equation there is
no term linear in the  quantum field $\chi$
in the action.

Dependence of the evolution operator in the external
field on the creation-annihilation fermion operators is completely
determined by \eq{s8}. Integral over quantum fluctuations produces the
determinant of the Schr\"odinger operator in the external field $\Phi$
(it is calculated in the Appendix \ref{chi}):
\begin{equation}
\log \left[ Det\Phi\left( T \right)\right] = -{1\over 4\pi } \int _0^T dt
 dt_1 \int _{-\infty}^\infty {dp\over 2\pi} \Phi \left( -p,t \right)
\Phi \left( p,t_1  \right) |p| \exp\left[ -i|p|v_f |t-t_1| \right] .
\label{s9} \end{equation}
The complete expression for the evolution operator in the external field
has the form:
\beq
 {\hat S \left( \Phi
\right)}=\exp{\left( {\cal S}_0 + \log\right[Det\Phi\left( T \right) \left]
\right)}|F><F|, \eeq

Now we  can express  the  evolution operator for the system of
interacting fermions in terms of this operator. We will use the
well-known identity \cite{Hub}:
\begin{eqnarray}
\exp  \left[ -{i\over2}\int_0^Tdt \int^\infty_{-\infty} {dp\over2\pi}
V \left( p \right)
\varrho \left( p,t \right) \varrho \left( -p,t  \right) \right]
=\nonumber\\
{1\over {\cal N }}\int {\cal D}\Phi \exp \left[ {i\over2}
\int_0^T dt \int^\infty _{-\infty}{dp\over2\pi} \Phi\left( p,t \right)
\Phi\left( -p,t \right) V^{-1} \left( p  \right) \right. \nonumber\\
-
\left.  {i\over2} \int_0^T dt \int^\infty _{-\infty}{dp\over2\pi}
\left( \varrho \left( p,t  \right)  \Phi\left( -p,t  \right) + \varrho
\left( -p,t  \right) \Phi\left( p,t  \right)\right) \right] .
\label{s10}
\end{eqnarray}
Here $V \left( p  \right)$ is the Fourier transform of the interacting
potential. The normalization coefficient $\cal N$ is
\begin{equation}
{\cal N } = \int {\cal D}\Phi \exp \left[  {i\over2} \int_0^T dt
\int^\infty _{-\infty}{dp\over2\pi}
\Phi\left( p,t \right) \Phi\left( -p,t \right) V^{-1} \left( p  \right)
\right] .
\label{s10a}
\end{equation}
In order to prove \eq{s10} it is sufficient to shift the variable of
integration $\Phi$ to $\Phi - V \varrho $ in the integral
\[
\int {\cal D}\Phi \exp \left[  {i\over2} Tr \left( \Phi^{\dag} \Phi V^{-1}
\right) \right] .
\]

Applying identity (\ref{s10}) to the functional integral that determines the
evolution operator for the Tomonaga-Luttinger model we express it in
terms of the evolution operator in the external field at the price of an
additional functional integration over the scalar field $\Phi(x,t)$:
\begin{equation}
{\hat S_{e-e}} = {1\over {\cal N } }
\int {\cal D}\Phi \exp \left[  {i\over2} \int_0^T dt \int^\infty _{-\infty}
{dp\over2\pi}
\Phi\left( p,t \right) \Phi\left( -p,t \right) V^{-1} \left( p  \right)
\right]  {\hat S \left( \Phi \right)}.
\label{s11}
\end{equation}

Expression (\ref{s11}) is explicit: while it is not possible to perform
the final integration in $\Phi(x,t)$ in closed form, it is
easy to obtain an arbitrary term of the evolution operator expanding it
in creation-annihilation operators. This will be enough in order to
calculate the evolution operator.

Indeed, let us expand the evolution operator in the external field in
powers $S_0^n$. The arbitrary term of the
expansion contains a number of Green functions in the external field
(\ref{s6a}) which are exponents linear in the external field. Together
with the action (\ref{s11}) and the determinant, \eq{s10a},
we get an integral of  Gaussian type in $\Phi(x,t)$
which can be easily performed. The result of the integration depends
of the electron-hole configuration considered.

Let us introduce the following system of notations  for the coordinates
entering the electron-hole creation-annihilation operators:
\begin{enumerate}
\item
We will denote by $x$ the coordinates of the right particles and by $y$ the
coordinates of the left ones.
\item
We will put the tilde on coordinates related to annihilation operators
(initial state) and leave coordinates of creation operators (final state)
without tilde.
\item
We will put primes on coordinates which are related to holes.
\end{enumerate}
It is convenient to proceed in the exponents of the Green functions
(\ref{s6a}) to  momentum space using the expression for the free
Feynman Green functions:
\beq G_{R,L}^0
\left( p, t, t_1 \right) =
\theta_{\pm p}\theta\left( t-t_1 \right) \exp \left[ \mp ipv_f \left( t
- t_1 \right) \right] - \theta_{\mp p}\theta\left( t_1-t \right) \exp
\left[ \mp ipv_f\left( t - t_1 \right)
\right]  .
\label{gr1a}
\eeq

Collecting all terms in the exponents arising from
the Green function we obtain the contribution to the action
linear in the external field $\Phi$:
\beq
{\cal S}_c = i\int_0^Tdt  \int _{-\infty}^\infty {dp\over 2\pi}\Phi
\left( -p,t \right) {\cal R}_c
\left( p,t \right) ,
\label{gr1}
\eeq
where the "current" ${\cal R}_c$ (which depends on the chosen
configuration) is equal to
\begin{equation}
{\cal R}_c\left( p,t \right) = {\cal R}_i\left( p \right)
\exp \left( -i|p|v_f t_1\right) +
{\cal R}_f\left( p \right)
\exp \left( -i|p|v_f \left( T-t_1 \right)\right),
\label{gr2}
\end{equation}
and
\begin{eqnarray}
{\cal R}_f\left( p \right) = \sum_{x..;x'..;y..;y'... } \theta
\left( p \right)\left[ \exp \left( ipx \right)
- \exp \left( ipx' \right) \right] +  \theta \left( -p \right)\left[
\exp \left( ipy \right)
- \exp \left( ipy' \right) \right] ,\nonumber \\
{\cal R}_i\left( p \right)= \sum_{\tilde x..;\tilde x'..;\tilde y..;
\tilde y'... }
\theta \left( -p \right)\left[ \exp \left( ip\tilde x \right)
- \exp \left( ip\tilde x' \right) \right] +  \theta \left( p \right)\left[
\exp \left( ip\tilde y \right)
- \exp \left( ip\tilde y' \right) \right]
\label{gr3}
\end{eqnarray}
for the initial (annihilation operators) and final (creation operators)
configurations respectively.
Coordinates $x,y$ in \eq{gr3} are the
coordinates of annihilation and creation operators for the
configuration we are interested in.
Finally, we get the following functional integral:
\[
\int {\cal D}\Phi \exp \left[ {i\over2} \int_0^T dt
dt_1 \int^\infty _{-\infty}{dp\over2\pi} \Phi\left( p,t \right)
\Phi\left( -p,t_1 \right) V^{-1} \left( p  \right) \delta \left( t-t_1
\right) \right.
\]\beq
- {1\over 4\pi } \int _0^T dt dt_1 \int
_{-\infty}^\infty {dp\over 2\pi} \Phi \left( -p,t  \right) \Phi \left(
 p,t_1  \right) |p| \exp\left[ -i|p|v_f |t-t_1|  \right] \\ +i\left.
\int_0^Tdt   \int _{-\infty}^\infty {dp\over 2\pi} \Phi \left( -p,t
\right) {\cal R}_c \left( p,t \right) \right]
\la{gr4}
\eeq
Here the first term is the action of \eq{s11}, the second is the
quantum determinant and the third comes from the Green
functions, \eq{gr1}.

The integral in \eq{gr4} is a Gaussian one: it can be calculated by
standard methods.  One has to find  the saddle point field
$\Phi_0$  and shift the variables of integration $\Phi \to\Phi -\Phi_0$.
The integral in fluctuations $\Phi-\Phi_0$ gives the shift of the
ground state energy due to the electron interaction and normalization
coefficient of the ground state wave function. We calculate
this integral in Appendix \ref{z}. The operator
structure of the evolution operator is determined completely by the
terms which appear as a result of substituting the saddle point
$\Phi_0$ in \eq{gr4}. We write them as an ``effective action":
\begin{equation}
{\cal S}_{eff}={i\over 2}\int^\infty
_{-\infty}{dp\over2\pi}\int_0^Tdt \Phi_0  \left( p, t  \right) {\cal R
}_c \left( p,t \right) .
\label{gr4a}
\end{equation}
The saddle point field $\Phi_0(x,t)$ obeys the integral equation:
\begin{equation}
{i\over V \left( p  \right)}\Phi_0  \left( p, t  \right) -
{1\over 2\pi } \int _0^T  dt_1
\Phi_0\left( p,t_1  \right) |p| \exp\left[ -i|p|v_f |t-t_1|  \right]
= -i {\cal R}_c \left( p, t  \right)
\label{gr5}
\end{equation}
which can be reduced to the following differential equation (to see this it is
sufficient to differentiate both sides of \eq{gr5} with respect to time):
\begin{equation}
\partial^2_t \Phi_0  \left( p, t  \right) +\omega_p^2
\Phi_0  \left( p, t \right)=0, \label{gr6}
\end{equation}
where
\begin{equation}
\omega_p = |p|v_f \sqrt{ 1 +{V\left( p  \right) \over\pi v_f}} .
\label{gr6a}
\end{equation}
The boundary conditions for this equation follow from the original integral
equation (\ref{gr5}):
\begin{eqnarray}
\partial_t \Phi_0  \left( p, 0  \right) -i|p|v_f\Phi_0  \left( p, 0  \right)
=
2i|p|v_fV\left( p \right) {\cal R }_i\left( p \right) \nonumber \\
\partial_t \Phi_0  \left( p, T  \right) +i|p|v_f\Phi_0  \left( p, T  \right) =
-2i|p|v_fV\left( p \right) {\cal R }_f\left( p \right) .
\label{gr7}
\end{eqnarray}
In the derivation of \eq{gr6} we have used the fact that our system is
electrically neutral and hence:
$${\cal R }_f\left( p=0,t\right)= {\cal R }_i\left( p=0,t \right) = 0 .
$$

The solution of the differential equation for the saddle point field  [\eq{gr6}]
gives:
\begin{eqnarray}
\Phi_0  \left( p, t  \right) = {-2|p|v_f V\left( p \right) \over
\left( \omega_p + |p|v_f  \right)
 \left( 1 - \xi_p^2  \right)}\left[  {\cal R }_i
\left( \exp \left( -i\omega_p   t  \right) +
\xi_p \exp \left( -i\omega_p  \left( T- t \right)  \right) \right)
\right.
\nonumber \\
\left. + {\cal R }_ f \left( \xi_p\exp \left( -i\omega_p  t  \right) +
 \exp \left( -i\omega_p  \left( T- t \right)  \right) \right)\right] ,
\label{gr8}
\end{eqnarray}
where
$$
\xi_p={1-\sqrt {1 +{V\left( p \right)\over \pi v_f}}\over 1+
\sqrt {1 +{V\left( p \right)\over \pi v_f}}}
\exp \left( -i\omega_pT \right) .
$$
Substituting the saddle point field into the expression for the effective
action (\ref{gr4a}) we obtain finally
\[
{\cal S}_{eff} = - {1\over L} \sum_{p\ne 0} {V\left( p \right) \over
1+\sqrt {1 +{V\left( p \right)\over \pi v_f}} }\cdot
{1 \over 1- \xi_p^2}\times
\]\beq
\times\left[ \left[{\cal R }_f \left( -p \right)
{\cal R }_f \left( p \right) +
{\cal R }_ i \left( -p \right) {\cal R }_i \left( p \right)\right] F_2
\left( p \right) +
2 F_1  \left( p \right) {\cal R }_f \left( -p \right) {\cal R }_ i
\left( p \right)\right] ,
\label{gr9}
\eeq
where we introduce the following two functions:
\begin{eqnarray}
F_1  \left( p \right) = {  \exp \left( -i|p|v_fT \right) -
\exp \left( -i\omega_p T \right) \over \omega_p
-|p|v_f  } +\xi_p {1 - \exp \left( -i\left( \omega_p +|p|v_f \right) T\right) \over \omega_p
+|p|v_f }\nonumber \\
F_2  \left( p \right) = {  1 - \exp \left(  -i\left( \omega_p +|p|v_f \right) T \right)
\over \omega_p
+|p|v_f  } +\xi_p {\exp \left( -i|p|v_fT \right) - \exp \left( -i \omega_p  T\right)
\over \omega_p
-|p|v_f } .
\label{gr10}
\end{eqnarray}

We return in the expression for the effective action (\ref{gr9}) to a
sum over the particle momentum $p_n=2\pi n/L$ according to the ordinary rule
\cite{bound}:
$$
\int^\infty _{-\infty}{dp\over2\pi} \longrightarrow
{1\over L} \sum_{p}.
$$
This will allow us to qualify different infrared divergences which appear
in the effective action.
Let us note that there is no term with $p=0$ in these sums. This fact
is related to the gauge invariance of the system: constant (in
space)  fields $\Phi(t)$ correspond to a pure gauge electrical
potential and should not contribute.

Let us proceed with the wave function of the ground state in
Tomonaga-Luttinger model. As  was mentioned above, in order to
separate the ground state we have to take the limit $T\to\infty$.
(This corresponds to the  case of zero temperature.)
We can omit oscillating exponents in this limit. As a result we
are left only with the function $F_2$ which is changed to
$$
F_2\left( p \right) \sim \left[  |p|v_f\left( 1+\sqrt {1 +{V\left( p \right)\over \pi v_f}}
 \, \right)\right]^{-1}.
$$
Effective action  factorizes into contribution of initial and final
states:
\[
{\cal S}_{eff}= - {1\over L}
 \sum_{p\ne 0} {V\left( p \right) \over |p|v_f \left[ 1+\sqrt {1
+{V\left( p \right)\over \pi v_f}}\,\right]^2 } \left[   {\cal R }_f
\left( -p \right) {\cal R }_f \left( p \right) + {\cal R }_ i \left( -p
\right) {\cal R }_i \left( p \right) \right]=
\]\beq
\hfill={\cal S}_{eff}^f +{\cal S}_{eff}^i .
\label{gr11}
\eeq

Besides $S_{eff}$ we have to calculate preexponential factors
which arise from  free Feynman Green functions. At $T\to\infty$
only Green functions with equal time arguments survive. As a
result, we see that the whole expression for the evolution
operator for large $T$ factorizes into the product of the wave
function of the ground state $|\Omega>$ and its complex conjugate
(see also below, section \ref{vo}). The final expression for the
wave function is of the form \beq | \Omega > = \sum_{n=0}^{\infty
}{1\over n!}\left[ \int {dxdx'\over2\pi i} { \hat a^{\dag
}_R\left( x \right) \hat b^{\dag }_R\left( x ' \right) \over x'-x
-i\delta } + \right. + \left. \int {dydy'\over2\pi i} { \hat
a^{\dag }_L\left( y  \right) \hat b^{\dag }_L\left( y ' \right)
\over y-y' -i\delta } \right]^n \exp {{\cal
S}_{eff}^f}\left(x,x',..,y,y',..  \right) |F > . \label{gr12} \eeq

Let us check, first of all, that the wave function of noninteracting
fermions ($V=0$) is $|F>$. The general term in the sum of \eq{gr12}
is a product of factors:
$$
\int {dxdx'\over2\pi i}
{ \hat a^{\dag }_R\left( x  \right) \hat b^{\dag }_R\left( x ' \right)
\over x'-x -i\delta } |F>.
$$
Let us note now that all singularities of the operator
$\hat b^{\dag }_R\left( x ' \right)$ are in the upper semiplane (see
definition \ref{abr}) as well as the pole of the Green function.
One can close the contour of $x^\prime$ in the lower semiplane and prove
that the corresponding integral vanishes. The only term which survives is
the term with $n=0$ and hence $|\Omega>=|F>$ as it should be
for non-interacting fermions.

A non-trivial answer for the wave function appears only owing to
singularities of the effective action. It is clear from the general
structure of the action (which is the product of $R_f(p)R_f(-p)$) that the
wave function contains only terms where both $R$- and $L$-particles are
present. All terms with only R (or only L) electrons or holes vanish.
The structure of the ground state wave function is a bit different
for different potentials. We will describe it below.

\subsection{Short range potential.}
\label{vo}
We begin with the short range potential:  $V\left( p \right) = V_0$.
The simplest possible contribution to the ground state  wave
function $| \Omega >$ (see \eq{gr12}) is
\begin{equation}
\int {dxdx'\over2\pi i}{dydy'\over 2\pi i}{ \hat a^{\dag }_R
\left( x  \right) \hat b^{\dag }_R\left( x ' \right) \over x'-x -i\delta }
{ \hat a^{\dag }_L\left( y  \right) \hat b^{\dag }_L\left( y ' \right)
\over y-y' -i\delta }  \exp
{{\cal S}_{eff}^f\left( x,x',y,y' \right) } .
\label{v1}
\end{equation}
The effective action ${\cal S}_{eff}$ for this term has the form:
\begin{eqnarray}
{\cal S}_{eff}^f\left( x,x',y,y' \right) = - {2\alpha\over L} \sum_{p_n> 0} {1 \over p_n}
\left[ \exp {ip_n\left( x-y +i\delta \right)}\right. + \nonumber \\
\left.\exp {ip_n\left( x'-y' +i\delta \right)} -
\exp {ip_n\left( x'-y -+i\delta \right)} -\exp {ip_n\left( x-y'
+i\delta \right)} \right] ,
\label{v2}
\end{eqnarray}
where
\begin{equation}
\alpha ={ V_0\over v_f\left[ 1+\sqrt {1 +V_0/ \pi v_f } \right]^2}.
\label{v3}
\end{equation}
The sums in \eq{v2} can be easily calculated. We obtain:
\begin{equation}
{\cal S}_{eff}^f\left( x,x',y,y' \right) =
{\alpha\over\pi}\log{{\left( x-y +i\delta \right)
\left( x'-y' +i\delta \right)\over
\left( x'-y +i\delta \right)\left( x-y' +i\delta \right)}}.
\label{v3a}
\end{equation}

Expression \ur{v1} describes the simplest possible complex in the
vacuum of the interacting fermions. This complex has all quantum
numbers equal to zero. In fact, it describes electron-electron
scattering (in the cross channel).
Correspondingly, all coordinates $x,x^\prime,y,y^\prime$
are close to each other. In general, this complex does not break
down any continuous symmetry.

However in the Tomonaga-Luttinger model a special situation arises.
The main contribution to the term \ur{v3a} comes from the region $x'-y;
x-y' \to 0$ (of order of the transverse size of the channel) but $x-y$
and $x^\prime-y^\prime$ can be arbitrarily large. In other words, the
complex decays into $R\bar L$ and $\bar R L$-pairs. As we shall see
such a wave function leads to a {\em spontaneous breakdown of chiral
symmetry}.

Let us consider first the limit of the strong interaction:
\begin{center}
\begin{equation}
{V_0\over \pi v_f} \gg 1 .
\label{v4}
\end{equation}
\end{center}
In this limit $\alpha /\pi \to 1$. It can be seen now that for
$\alpha/\pi=1$ the poles ($x=x^\prime$ and $y=y^\prime$)
corresponding to free fermions  are cancelled
completely by the fermion-fermion interaction (described by
$\exp(S_{eff})$) with the effective action of \eq{v3a}. Instead we
obtain new poles in the points $x^\prime=y-i\delta$ and
$y^\prime=x+i\delta$. Recalling that $\hat b^{\dag }_R\left(
x^\prime \right)$ is analytical in the lower and $\hat
b^{\dag }_L\left( y' \right)$ in the upper semiplane we can integrate further
over $x^\prime$  and $y^\prime$. As a result we get the following
contribution to the  ground  state wave function:
\begin{equation}
\int dx \hat a^{\dag }_R \left( x  \right)
\hat b^{\dag }_L\left( x  \right) \int dy \hat a^{\dag }_L\left( y  \right)
\hat b^{\dag }_R\left( y \right) .
\label{v5}
\end{equation}
Thus the 4-particle complex decays into 2 non-interacting
"bosons". They are neutral in the electric charge but have a {\em
non-zero chirality} $\pm 2$.

One can check that no other connected complexes appear in the
limit of strong interaction. Let us consider, for example,  charged
complexes. The four fermion contribution is exhausted by \eq{v5},
so we have to consider a 6-fermion complex:
\begin{eqnarray}
{ \hat a^{\dag }_R\left( x  \right)
\hat b^{\dag }_R\left( x ' \right) \over x'-x -i\delta } { \hat a^{\dag
 }_R \left( x _1 \right) \hat b^{\dag }_R\left( x ' _1\right) \over
x'_1-x_1 -i\delta } { \hat a^{\dag }_L\left( y  \right) \hat b^{\dag
}_L \left( y ' \right) \over y-y' -i\delta }  \nonumber\\ {\left( x -y
+i\delta \right) \left( x_1 -y +i\delta \right) \left( x' -y' +i\delta
\right) \left( x '_1-y' +i\delta \right) \over \left( x -y' +i\delta
\right) \left( x _1-y' +i\delta \right) \left( x' -y +i\delta \right)
\left( x'_1-y +i\delta \right)} .
\label{v6}
\end{eqnarray}
This complex, indeed, decays into 2 fermions with
$x_1\to y'\to x$ and $x'\to y\to x'_1$  (relative distance
$x-x^\prime$ supposed to be large). These fermions are of the
form $\hat a^{\dag }_R\left( x  \right) \hat a^{\dag }_R\left( x  \right)
\hat b^{\dag }_L\left( x  \right)$ and
$\hat a^{\dag }_L\left( x'  \right)\hat b^{\dag }_R\left( x ' \right)
\hat b^{\dag }_R\left( x ' \right)$. Hence this contribution is
zero owing to the Pauli principle.
One can consider also more complicated
configurations which could produce charged connected complexes and
check that they do not appear in the wave function of the ground state.

The Pauli principle allows one more complex which describe
scattering of chiral pairs:
\begin{center}
$\hat a^{\dag }_R\left( x  \right) \hat b^{\dag }_L\left( x  \right)\hat a^{\dag }_L
\left( x  \right)
\hat b^{\dag }_R\left( x  \right) .$
\end{center}
One can extract the corresponding contribution from the
connected part of the general expression \eq{v1}. The integral over
$x'$ and $y'$ is easily calculated and we obtain:
\begin{center}
$
\int dx dy\hat a^{\dag }_R\left( x  \right) \hat b^{\dag }_L\left( x  \right)\hat
a^{\dag }_L\left( y  \right)\hat b^{\dag }_R\left( y  \right)
\Phi \left( x-y  \right),
$
\end{center}
where
$$
\Phi \left( x- y  \right) ={ -i\delta \over y-x -2i\delta} \left( 2 +  {i\delta
\over y-x -2i\delta}\right) .
$$
The function $\Phi \left( x- y  \right)$ is finite at any $x$, $y$
(even in the point $x=y$) so its contribution to the integral vanishes
in the limit $\delta\to +0$. In other words, in the limit of infinitely
strong interaction the chiral pairs do not interact. This interaction
appears, however, in the next approximations in the inverse coupling
constant (see Section \ref{K}).

To obtain the complete expression for the ground state wave
function we have to consider complexes with 8,12\ldots particles
and separate the connected parts out of these complexes. this  is not
necessary, however, since, according to a general theorem
\cite{FS}, the complete wave function is the exponent of the connected
complexes \cite{log} and we have proved that the only
connected complexes are the chiral pairs (\ref{v5}). On the other hand
the total chirality $C$ of $|\Omega>$ should be zero and only
terms with $C=0$ can appear in the expansion of $|\Omega>$. To
take this into account  we introduce the projector $P_{C=0}$ onto
the state with chirality zero. Then the wave function can be
written as
\begin{eqnarray}
| \Omega > = \sqrt {Z_0}\hat P_{C=0}\exp \left[ \int dx \hat a^{\dag }_R
\left( x  \right) \hat b^{\dag }_L\left( x  \right) +
\int dy \hat a^{\dag }_L\left( y  \right)
\hat b^{\dag }_R\left( y  \right)
\right] | F >
\label{v7}
\end{eqnarray}
The normalization coefficient $Z_0$ is  calculated in  Appendix \ref{z}.
We have already discussed the wave function in  Section \ref{r1}. It corresponds
to an unbroken symmetry phase in spite of the presence of an infinite
number of chiral pairs with zero momentum. If the chiral symmetry
is broken, the states with different chirality should be {\em
degenerate} in energy. This is not the case if the size of system is
finite --- the energy of the state with $C=0$ is still minimal and
order parameter $\Delta$ is zero.

Wave function (\ref{v7}) corresponds to the state with minimal
possible energy. Hence it is the wave function of the system
at $\Theta=0$. Turning to $\Theta\neq 0$ let us note that there
are two types of exponentials (see below \eq{t2}): first with
$\exp(-v_fp/\Theta)$ and the second is $\exp(-v^c_fp/\Theta)$
(with renormalized
Fermi velocity). Corrections of the second type correspond to
excitations and we will omit them. However there are no excitations
with $\omega_p=v_fp$ (it can be seen, e.g., by method of bosonization).
In fact, these exponentials describe the change
of the ground state with temperature \cite{matrix}. Obviously for
$ \Theta\gg 2\pi v_f/ L$ the exponential factor is
not small but the preexponential factor, i.e. the Green function with imaginary
time differences about $1/\Theta$, gives the  smallness. The latter is
compensated by the action because it is proportional to $\log {1/\Theta}.$
This  is the case for the temperature region under consideration.  In the opposite
case the Green function (\ref{s6}) is not valid. One
can use \eq{gr1a} but it is impossible to transform the sums over
$p_n$ to integrals in order to get  \eq{s6}. As a
result, the  Green function will be proportional to a small
exponential factor. It can not be compensated by the logarithmic
divergence from action and the whole term will be small. Therefore the
ground state wave function (\ref{v7}) is valid provided
\begin{eqnarray}
\Theta\ll \Theta_{chiral}= {2\pi v_f\over L}
\label{t1}
\end{eqnarray}
Of course one assumes that the number of states  is large , i.e. $p_fL\gg 1$.
This allows a transition from the sums
to integrals in the expressions independent of $\Theta$ (or $T$).

In the region of higher temperatures,
$
 \Theta_{chiral}\ll \Theta,
$
\eq{s6} for  the Green function is applicable.
In this case
after same algebraic transformations effective action ${\cal S}_{eff}$
(\ref{gr9}) can be rewritten in the form of
$$
{\cal S}_{eff}=-{\pi\over L}\sum_{n\ne 0}{1\over |p_n|}
\tanh{|p_n|v_f^c\over 2\Theta}
\left[   {\cal R }_f \left( -p \right) {\cal R }_f \left( p \right) +
{\cal R }_ i \left( -p \right) {\cal R }_i \left( p \right)\right]
$$
\begin{equation}
-{2\pi\over L}\sum_{n\ne 0}{1\over |p_n|}{\exp{-|p_n|v_f\over \Theta}
- \exp{-|p_n|v_f^c\over \Theta}
\over 1+\exp{-|p_n|v_f^c\over \Theta}}{\cal R }_f
\left( p \right){\cal R }_i \left(- p \right) ,
\label{t2}
\end{equation}
where  $v_f^c =v_f \sqrt {1 +V_0/ \pi v_f}.$
If
\begin{equation}
\Theta_{chiral}\ll \Theta\ll 2\pi v_f^c/L
\label{t4b}
\end{equation}
than equation  (\ref{t2}) can be transformed to
$$
{\cal S}_{eff}=-{\pi\over L}\sum_{n\ne 0}{1\over |p_n|}
\left[   {\cal R }_f \left( -p \right) {\cal R }_f \left( p \right) +
{\cal R }_ i \left( -p \right) {\cal R }_i \left( p \right)\right]
$$
\begin{equation}
-{2\pi\over L}\sum_{n\ne 0}{1\over |p_n|}\exp{-|p_n|v_f\over \Theta}
{\cal R }_f \left( -p \right) {\cal R }_i \left( p \right) ,
\label{t3}
\end{equation}
\cite{rem2a}.
 So, in the temperature region  of interest one should take
into account another 4-fermions contribution to the ground state:
\begin{equation}
\int {dxd\tilde x\over2\pi i}{dy'd\tilde y'\over 2\pi i}{ \hat a^{\dag }_R
\left( x  \right) \hat a_R\left( \tilde x  \right) \over \tilde x - x +v_fT
-i\delta }
{ \hat b^{\dag }_L\left( y'  \right) \hat b_L\left( \tilde y ' \right)
\over \tilde y' - y' -v_fT+i\delta }  \exp
{{\cal S}_{eff}^f\left( x,\tilde x,y', \tilde y' \right) }.
\label{t4}
\end{equation}
At lower temperature this contribution is exponentially small.
The action for this configuration is
\begin{equation}
\log{{\left( \tilde y' - y' -v_fT+i\delta \right)\left( x- \tilde
x -v_fT +i\delta \right) \over \left( x -y'+i\delta \right) \left(
\tilde y' -\tilde x +i\delta \right)}}. \label{t4a}
\end{equation}

Thus one has a similar result: a pair $\hat a^{\dag }_R\left( x  \right)
\hat b^{\dag }_L\left( x  \right)$ in $|\Omega>$ and $\hat a_R\left(
\tilde x  \right)
\hat b_L\left( \tilde x  \right) -$  in $<\Omega|.$ However the existence
of an extra
pair implies that the whole chirality $C$ of the state is nonzero.
So the states with any
$C$  exist. Their energies differ by  values of the order of $2\pi v_f/L .$
At the temperature \eq{t4b}
these states can be considered as degenerate. Then a state with fixed chirality
is unstable relative to an infinitesimal interaction which is mixing right and
left particles (e.g. infinitesimal back scattering). Similar to the
theory of superconductivity the real ground state of the system is
a mixture of states with different chirality, \eq{rt5}, but a fixed chiral
phase $\theta$.

In order to  prove that
\begin{equation}
\Theta_c = \omega\left(2\pi /L\right)=2\pi v_f^c/L
\label{t6}
\end{equation}
is the phase transition
temperature one should consider
the higher temperature region:
$
\Theta \gg \Theta_c
$.
The logarithmic contribution to the action  ${\cal S}_{eff}$ arises from
$n\gg n_{min}\sim L\Theta/2\pi v_f^c \gg 1.$ At
smaller $n$ the logarithmic divergence
does not exist. So
$$
{\cal S}_{eff}=-{\pi\over L}\sum_{|n|>n_{min} }{1\over |p_n|}
\left[   {\cal R }_f \left( -p \right) {\cal R }_f \left( p \right) +
{\cal R }_ i \left( -p \right) {\cal R }_i \left( p \right)\right]
$$
\begin{eqnarray}
-{2\pi\over L}\sum_{|n|>n_{min}}{1\over |p_n|}\exp{-|p_n|v_f\over \Theta}
{\cal R }_f \left( -p \right) {\cal R }_i \left( p \right)
\label{t7}
\end{eqnarray}
One can compare it with \eq{t2}. The sums  in \eq{t7} are
calculated in Appendix \ref{sum}. As a result the logarithms from
 \eq{v3a} have to be replaced by
\begin{equation}
\log{{\left( \Delta x+i\delta \right) \over \left( \Delta
x'+i\delta \right)}} \to \int^{-\left( \Delta x +i\delta
\right)}_{-\left( \Delta x'+i\delta \right)} {dz\over z} \exp
\left( -{iz\over \zeta \left( \Theta \right)} \right) ,
\label{t8a}
\end{equation}
where
\begin{equation}
\zeta  \left( \Theta \right) = v_f ^c /\left( \Theta - \Theta_c \right)
\label{t8a1}
\end{equation}
 is the coherence length.

The r.h.side of \eq{t8a} can be expressed by an exponential integral
exponent function with imaginary argument. In order to prove that
$\zeta  \left( \Theta \right)$ is the coherence length one should
notice that at the length $\Delta x \ll \zeta  \left( \Theta
\right)$ the  r.h.s. of \eq{t8a} tends to $\log{\left(\left( \Delta x+i\delta
\right) / \left( \Delta x'+i\delta \right)\right)}, $ i.e. in such
case the system is characterized by the wave function \eq{rt5}.
(Indeed, in this case it is possible to repeat the calculations
made in Section (\ref{vo}) if one separates all connected
complexes by the distances smaller than $\zeta  \left( \Theta
\right)$.) Thus in a region of a specimen smaller than $\zeta $
one has a coherent state. In the opposite case (distances
between pairs, $\Delta x = |x-y|$, are larger than $\zeta  \left(
\Theta \right)$) expression under the integrand begins to oscillate
and the divergence does not exist. As a result, one will have
small corrections to the action of about
$$\exp\left( -i\Delta x/\zeta  \left( \Theta \right) \right)
\zeta  \left( \Theta \right)/\Delta x  .$$
So  the 4-fermion contribution (\ref{t4}) leads to  a term
\begin{equation}
\int dxdy   \left( { \zeta \left( \Theta \right)\over |x-y| } \right)^2\hat a^{\dag }_R
\left( x  \right) \hat b^{\dag }_L\left( x  \right)|F>
<F| \hat b_L\left( y  \right) \hat a_R\left( y  \right)
\label{t8b}
\end{equation}
in the evolution operator. Thus at the distance $|x-y| \gg\zeta$
we have configurations with free bosons. Consequently, at such
scale the state is noncoherent.  Therefore a long range order does not
exist at lengths  larger than $\zeta  \left( \Theta \right)$.
One can also check this directly. Let us calculate the  contribution
of the state (\ref{t8b}) to the order parameter density
correlator: $<|\hat a^{\dag }_R \left( y_1  \right) \hat b^{\dag
}_L\left( y_1  \right)\hat b_L\left( x _1 \right) \hat a_R\left( x
_1 \right)|>.$ After integration over $x_1$ and $y_1$ one has the
contribution of this state to $\Delta^2$:
$$
\int^{L/2}_{-L/2} dx \left(  {\zeta \left( \Theta \right)^2\over L }
\right) \sim \zeta^2 \left( \Theta \right) .
$$
As $\Delta$ does not depend on $L$ one has the normal phase (see Section
\ref {r1}) with low symmetry phase fluctuations. This means
that $\Theta_c$ is indeed  the phase transition temperature and $\zeta$
is  the coherence
length. Besides, one has  more obvious definition  $\Theta_c$:
$$
\zeta \left( \Theta_c \right)\sim L .
$$
In this case the whole system
can be described by the broken symmetry wave function (\ref{rt5}). Hence one should think
that the low symmetry phase is realized if $\Theta< \Theta_c$.
The above discussion should make it clear  that
this transition is smeared over the temperature region about $\Theta_c$
as it should be for the finite size specimen.

\subsubsection{Kosterlitz-Thouless phase.}
\label{K}

Let us prove that a Kosterlitz-Thouless phase \cite{KT}  is likely to form
in the Tomonaga-Luttinger model if one takes into account corrections to
the action due to $ \pi v_f/ V_0 $.

We  begin with the case of zero temperature and consider again
the 4-fermion contribution  \eq{v1} to the ground wave function
\begin{equation}
\int {dxdx'\over2\pi i}{dydy'\over 2\pi i}{ \hat a^{\dag }_R
\left( x  \right) \hat b^{\dag }_R\left( x ' \right) \over x'-x -i\delta }
{ \hat a^{\dag }_L\left( y  \right) \hat b^{\dag }_L\left( y ' \right)
\over y-y' -i\delta }
 \left[ {\left(  x -y +i\delta\right) \left(  x' - y' +i\delta\right) \over
\left(  x' -y +i\delta\right) \left( x - y'
+i\delta\right)}\right]^{\alpha_0}|F>, \label{v8}
\end{equation}
where $\alpha_0\equiv \alpha / \pi$.  For simplicity we will
consider $\alpha_0$ close to unity.  Let us consider the
configuration with two connected chiral complexes separated by
a distance $R$ large compared to the transverse size of the
channel $d$: $x'-y; x-y'\sim d \to 0$ and $|x-x'| \sim R $ ,
$|y-y'| \sim R \to \infty $.  The contribution we are interested in
is determined by
two cuts: $y'=x+ i\delta $ and $x'=y+i\delta $ and is proportional to
$$ \left( 1-e^{2\pi i
\alpha_0} \right) \int^x_{-\infty}{dy'\over 2\pi i}\hat b^{\dag
}_L\left( y ' \right) {1\over \left(y' -
x\right)^{\alpha_0}}{\left(x'-y' \right)^{\alpha_0} \over
\left(y'-y \right)} .
$$
The last factor of the integrand is of order the $1/R^{1-\alpha_0 }$. Distances inside the
pair $y^\prime-x$, $x^\prime-y$ are of order of $d$. The
contribution of the distant chiral pairs to the integral (\ref{v8}) is
\begin{equation}
\int dxdy \hat a^{\dag }_R \left( x  \right) \hat b^{\dag
}_L\left( x \right)  \hat a^{\dag }_L\left( y  \right) \hat
b^{\dag }_R\left( y \right) \left( {d\over |x-y| }
\right)^{2\left( 1-\alpha_0 \right)}|F>.
\label{v9}
\end{equation}
In the temperature region (\ref{rt1}) we can consider also contributions
of the states with $C\neq 0$ to the ground state. The simplest
contribution comes again from \eq{t4} and has the form
\begin{equation}
\int dxdy   \left( { d\over |x-y| } \right)^{2\left( 1- \alpha_0
\right)} \hat a^{\dag }_R \left( x  \right) \hat b^{\dag }_L\left(
x  \right)|F><F| \hat b_L \left( y  \right) \hat a_R\left( y
\right) \label{rv9}
\end{equation}
As can be seen from \eq{v9} and \eq{rv9} the probability to find
chiral pairs at the distance $R$ is $P(R)=|\Phi(R)|^2\sim
1/R^{2(1-\alpha_0)}$\cite{rem4}.  This probability decreases with
$R$ but much more slowly than in the theory without interaction. The
average distance between pairs
$$<R>=\int_0^L\!\!\! dR \;R\; P(R) \sim L^{2\alpha_0} $$
diverges as $L\to\infty$.  It is instructive to consider the same
quantities in the theory with non-interacting electrons. Here the
probability to find a chiral pair is $P_{free}=\left( { d\over
|x-y| } \right)^2$ (see \eq{t8b}). As we have seen it results in
the independence of $\Delta$ on $L$. The other limiting case is the
system with non-zero density of the order parameter. Here the
probability to find a chiral pair does not depend on the distance
$R$ at all while $\Delta\propto L $. The probability under discussion
has an intermediate behavior. As a result  $\Delta$ will increase
with $L$ but the power will be smaller than unity.
Both these properties can be
considered as a definition of Kosterlitz-Thouless phase.

The temperature of the phase transition $\Theta_c$ in the
Kosterlitz-Thouless system, at $\alpha_0<1$, to the unbroken phase is
of the same order as in the limit of the infinitely strong
interaction. Indeed our estimate of $\Theta_c$ in section \ref{vo}
was based on the logarithmic divergence of the action. This
divergence exists also at $\alpha_0<1$ and hence our expressions
for $\Theta_c$  and correlation length $\zeta$ are valid in this
case too.

The wave function for the Kosterlitz-Thouless phase does not
have the simple form of \eq{rt5} since the interaction of chiral
pairs is non-zero. Also chiral complexes with more than two
particles are present in the wave function of the ground state.
However properties of this phase are quite similar to properties
of the phase with broken symmetry which appears in the limit
of infinitely strong interaction.

\subsection{Cooper channel.}
\label{cup}

Let us discuss in brief the case of the  attractive  short
range potential $V=V_0\delta(x)$, $V_0<0$ (Gorkov's potential
\cite{L1X}). This potential can be used only if the interaction is
sufficiently weak:
\begin{equation}
{ |V_0| \over\pi v_f} < 1, \label{cup1}
\end{equation}
otherwise the spectrum of excitations
acquires an imaginary part (see \eq{gr6a}).
This demonstrates that the system of
one dimensional electrons tends to collapse in this case and the
point-like potential has to be modified. This effect is similar to
the well-known instability in the system of interacting oscillators
\cite{GP}.

If inequality (\ref{cup1}) is fulfilled we can use \eq{v8}  derived
above but with  $\alpha_0 <0$. In this case the system is in the
Kosterlitz-Thouless phase too. To prove this statement we consider
again four-fermion contribution to the wave function
 with two Cooper pairs separated by distance $R$
($x-y; x'-y' \to 0,$ $|x-x'| \sim |y-y'| \sim R \sim L$).
The corresponding
contribution to the wave function is of the form:
\[
\hat a^{\dag }_R \left( x  \right) \hat a^{\dag
}_L\left( x  \right)\hat b^{\dag }_R\left( y' \right) \hat b^{\dag
 }_L\left( y' \right) ,
\]
and the probability to find such a configuration behaves as
$R^{2\left( 1-|\alpha_0|\right)}$, i.e. it decays more slowly than in the
theory without interaction.

\subsection{Coulomb interaction.}
\label{cul}

 We are interested in the long range Coulomb potential also in the
limit  of strong interaction. However as was already mentioned
in Section \ref{G},since the length of the three-dimensional screening
should be still larger than the size of the channel, we must have
$p_fa_b \gg 1$ \cite{Bohr}. This means that the parameter
related to the strength of the Coulomb interaction should satisfy
the inequality:
\begin{equation}
{V\left( p  \right)\over \pi v_f}= {2\over \pi
p_fa_b}\log{2p_f\over p} \gg 1 . \label{cul1}
\end{equation}
This can be large only due to a large logarithm. The argument of this
logarithm  is of order $p_fL/ \pi\gg 1$. For this reason the integrals
over momentum in the expression for the action ${\cal S}_{eff}$, \eq{gr11},
should be cut at
\[
p_{max} \sim 2p_f \exp \left(
{-\pi p_fa_b\over 2}  \right) \ll p_f
\]
This cut-off is not essential if $p_{max}L \gg 1.$ Then every term of
the action can be written as
\begin{eqnarray}
-\int^{\infty }_{1/L} {dp \over p} \exp{ \left( ip \delta x
\right) } /  \left( 1 + 3/2 \sqrt {\pi v_f/e^2}  \log^{-1/2}
\left( 2p_f/p  \right) \right) ^2    \sim \nonumber \\ \log \delta
x - 6 \sqrt {\pi v_f/e^2}  \log^{1/2}\left( 2p_f\delta x  \right)
\nonumber .
\end{eqnarray}
Separating the four-fermion contribution in two chiral pairs
$ \hat
a^{\dag }_R \left( x  \right) \hat b^{\dag }_L\left( x  \right)  \hat
a^{\dag }_L\left( y  \right) \hat b^{\dag }_R\left( y  \right) ,$
we find that the probability to find the chiral pair at a large
distance $|x-y|$ from its counterpart behaves like
\[
\exp \left( - 12\sqrt {\pi v_f/e^2}  \log^{1/2} \left( 2p_f  |x-y|
\right) \right) .
\]
It decays more slowly than any power of $|x-y|$ \cite{rem5}. Strictly
speaking, this behavior does not correspond to the Kosterlitz-Thouless
phase  but it is clear that physically these two phases are quite
similar.  Let us note also that in the formal limit $\left( \pi
v_f/e^2 \right) \log \left( 2p_f L  \right) \to 0$ we have again
the condensate of independent chiral pairs with the wave function
of \eq{rt5}.

\begin{acknowledgments}
We are grateful to V.L.Gurevich, Yu.M.Galperin and V.I.Kozub for a number of interesting
discussions and V.L.Gurevich, W.v.Schlippe for reading the manuscript.
V.V.A. also acknowledge a partial
support for this work by Grant ${\cal N}^0$~N-Sh.2242.2003.02.
\end{acknowledgments}

\appendix
\section{ Evolution operator for  systems of fermions.}
\label{ev}

In this appendix we derive the representation for the evolution
operator of fermions in the external field as a functional
integral with definite boundary conditions.

In the Schr\"odinger representation the evolution operator $S(T)$
is
$$
S[T]=T\exp(-i \int_0^T H dt)|F><F| ,
$$
where $H$ is the fermion Hamiltonian in the external field which
is bilinear in the fermion fields.  As we have seen the general
problem with electron- electron interaction can be reduced to this
problem at the price of the integration over the external field. For
simplicity we begin with a model with an empty ground state
$|0>$ rather than the Fermi one. (This allows us to write the
equations in a more compact form.) Besides we will omit the spatial
arguments.

Let us  divide the time interval $T$  in $N$ infinitesimal pieces
$\delta = T/N$ (with the point $i=N$ corresponding to $t=0$ and
$i=1$ to $t=T$) and introduce the sum over the complete set of
quantum mechanical states $|k><k|$ in the intermediate points:
\begin{equation}
S[T]= \sum_{n_i}|k_N><k_N|\left(1-i\delta H\right)|k_{N-1}>
\ldots<k_2|\left(1-i\delta H\right)|k_{1}><k_1| .
\label{a2}
\end{equation}

For any complete set of wave functions in the secondary quantization
representation we have:
\begin{eqnarray}
\sum_n|k_i(n)> <k_i(n)| = \int
{\cal D}\xi_i{\cal D}\xi^{\dag}_i\exp\left(-Tr\xi^{\dag}_i \xi_i\right)
\nonumber\\ \exp\left(Tr  \xi_i \hat a^{\dag} \right)|0><0|\exp\left(-Tr
\xi_i ^{\dag}\hat a \right) , \label{a1}
\end{eqnarray}
(index $n$ corresponds to the set of all quantum numbers). The
Grassmann variables $\xi$ are defined in the usual way:
\[
\int
d\xi_i(n)=0, \int d\xi_i(n)\xi_i(n)=1,  \left[ \xi_i ^{\dag}(n),
\xi_i(n) \right]_+ = 0 ., {\cal D}\xi_i =\prod_n d\xi_i(n) .
\]
Eq.(\ref{a1}) can be proved by direct comparison of the left and
right-hand sides. We use this representation to rewrite the sum over
states as a functional integral.

At every point $i$ we obtain the following matrix element of the
Hamiltonian:
\begin{eqnarray}
\exp\left(-Tr \xi^{\dag}_i
\xi_{i}\right)<0|\exp\left(-Tr  \xi_i ^{\dag}\hat a \right)
\left(1-i\delta H\left(\hat a^{\dag} ,\hat a\right)\right)
\exp\left(-Tr \xi^{\dag}_{i +1}\xi_{i+1}\right)\\
\exp\left(Tr \xi_{i+1} \hat a^{\dag} \right)|0>\nonumber .
\label{a2a}
\end{eqnarray}
To calculate this matrix element we move all
creation operators to the right. For the Hamiltonian $H$
depending linearly on $\hat a,\hat a^+$  e.g. for the Hamiltonian in the
external field the result is
$$
\exp\left( Tr\xi^{\dag}_i\left(\xi_{i+1}-\xi_{i}\right) +i\delta Tr H \left( \xi^{\dag}_i,
\xi_{i+1}\right)\right) .
$$
Thus the result of the calculation is that  creation-annihilation
operators in the Hamiltonian are substituted by Grassmann
variables $\xi,\xi^+$.

The product over all intermediate points at $N\to\infty$ tends to
\[
\exp(-\int_0^T\!\!\! dt\: \overline{\Psi} \left( t\right)\left[
\partial _t + i{\cal H} \right]\Psi \left( t\right))=
\exp{i\int_0^T\!\!\!dt\: {\cal L}} ,
\quad
{\cal L}=\overline{\Psi}\left[i\partial _t-{\cal H}\right]\Psi
\]
where ${\cal L}$ is the Lagrangian of the system. This expression
should be integrated over variables $\Psi,\bar{\Psi}$ at all
intermediate points in time. The boundary points are specific,
however. Creation operators entering $|k_N>$ and annihilation operators
entering $<k_1|$ are not contracted. They are variables on which the
evolution operator depends.

Let us integrate over all intermediate variables and consider the
answer as a function of Grassmann   variable $\xi_1^+$ (and
$\xi_N^\dag$).  This function can be only linear one: $A+B\xi_1$.
Then the last integration in $\xi_1,\xi^+_1$ is of the form:
\[
\int {\cal
D}\xi_1{\cal D}\xi^{\dag}_1\exp\left(-Tr \xi^{\dag}_1 \xi_1\right)
\exp\left( -Tr  \xi_1^{\dag}\hat a \right)\left( A_1+Tr B_1\xi_1 \right)
= A_1+Tr B_1 a
\]
Thus, we see that the variable $\xi_1$ should be substituted by
an annihilation operator. Integrating over $\xi^+_N$ we conclude
that $\xi^+_N$ is substituted by a creation operator.

Finally, we can formulate the following recipe: to calculate the
evolution operator one has to integrate $\exp(i\int_0^T{\cal L})$
over $\Psi,\bar{\Psi}$ at all intermediate points. At $t=0$
the value of $\Psi$ is fixed to $\hat a$, at $t=T$ $\bar{\Psi}$
is fixed to $\hat a^+$. The values of $\bar{\Psi}$ at $t=0$ and $\Psi$ at
$t=T$ remain arbitrary.  As a result, operators $\hat a$ and $\hat a^+$
are defined at different times. Therefore one should consider
theirs  here as anticommutating.

If the ground state of our system is a filled Fermi sphere we
have to introduce two type of creation-annihilation operators
$\hat a^\pm,\hat b^\pm$ corresponding to electrons and holes.
Then we can apply
the above derivation to this case as well.
One should introduce negative ($\Psi^-$) and positive
($\Psi^+$) frequency parts of $\Psi$ variables and double the number of
variables $\xi$.

\section{ Calculation of $Det \Phi .$ }
\label{chi}

We will calculate the functional integral over the fields $\chi$ and $\bar {\chi}$. They obey
zero initial conditions.
\begin{equation}
Det \Phi = \int {\cal D}\overline{\chi}{\cal D}\chi \exp\left( i\int^T_0
dt\int dx\overline{\chi}
\left( i\partial_t - {\cal H}_{ext}\left( x\right) \right)\chi\right) ,
\label{c1}
\end{equation}
where ${\cal H}_{ext}={\cal H}_0\left( x\right) + \Phi \left( x,t\right) .$
In an ordinary case $Det\Phi $ can be calculated in the usual way using
the identity
$$ \log \left[ Det \Phi  \right] = Tr \log \Phi .  $$
After differentiation over $\lambda$ the right-hand
side of this identity is represented  as
$$ Sp\left[ -i \int_0^1d{\lambda} \left( -\partial_t
- i{\cal H}_0\left( x\right) - i\lambda \Phi \left(
x\right)\right)^{-1} \Phi \left( x \right) \right] , $$
where the inverse operator is the Green function with the same
arguments.  Usually
the result does not depend on the order of arguments. However in the
theory with Adler-Schwinger anomaly the sequence of time and spatial
arguments is essential.  The simplest way is to assume to
make spatial arguments equal first.
In this case the result will be
in contradiction with gauge invariance of the theory.
In the paper \cite{P} the procedure
free of  this difficulty was suggested. It does not exist in this case
because all calculations
are done with non-equal variables up to the end.
The procedure is based on the Heisenberg equation
for the electron evolution operator
$ \hat S \left( \Phi \right) =
Det\Phi\exp {{\cal S}_0\left( T\right)}$
at the external field (without direct electron-electron
interaction). In Heisenberg representation one has
$$
i{\partial \hat S \over \partial T} = \left[ H_{ext}, \hat S \right] ,
$$
where $ H_{ext}$ is the noninteracting electrons Hamiltonian  (the external field
is dependent on the time $T$), and the action ${\cal S}_0 $ is defined by the \eq{s8}.
One should note that all creation operators are defined at the moment $T$ and the
annihilation operators   at $t=0$, therefore in the commutator $ \left[ H_{ext}, {\cal
S}_0\left( T\right) \right]$ only the terms with creation operators
do not commute with ${\cal S}_0 $.

One can rewrite the latter equation in the following form:
\begin{equation}
i{\partial \log {Det\Phi} \over \partial T} = \exp{ \left( -{\cal
S}_0 \right) }\left[ H_{ext}, \exp{{\cal S}_0} \right] -i{\partial
{\cal S }_0\over \partial T} \label{det1}
\end{equation}
In order to calculate  the commutator in this equation  one can use the well-known identity:
\begin{eqnarray}
\left[ \hat a\left( x\right) , \exp{\left( \int dx' K\left( x'
\right)\hat a^{\dag} \left( x' \right)\right)} \right] \nonumber \\=
\int dx_1 \Delta \left( x_1-x' \right) K\left( x_1 \right)
\exp{\left( \int dx' K\left( x' \right) \hat a^{\dag}\left( x'
\right)\right)} \nonumber,
\end{eqnarray}
which  can be proved by expanding the exponential functions.
 (Here $K$  is   anticommutating with $\hat a$ operator and
$\Delta \left( x_1-x' \right)$ - is the anticommutator $\{ \hat a\left( x\right) ,
\hat a^{\dag}\left( x' \right)\}$ is defined by \eq{g10}).
The left-hand side of  \eq{det1} is a c-number; this means that all operators from
right-hand side  of
this equation have to vanish. The c-number  parts arise only from the
following commutators:
\begin{eqnarray}
\int dx \Phi \left( x \right) \left[\hat b \left( x\right) \hat a\left(
x\right) , \exp{\left( \int dy dy'
\hat a^{\dag}\left( y' \right) G\left(  y'T,yT- \varepsilon \right)
\hat b^{\dag}\left( y \right)\right) } \right] \nonumber .
\end{eqnarray}
As a result one has
\begin{eqnarray}
i{\partial \log {Det\Phi} \over \partial T} = \int{dx dy dy'\over
\left( 2\pi i \right)^2}\Phi \left( x,T \right)
\nonumber \\
\left[  {G_R\left( y'T,y T- \varepsilon \right) \over \left( y'- x -i\delta \right)
\left( y- x -i\delta \right)} + {G_L \left( y'T,y T- \varepsilon \right)
 \over \left(  x-y' -i\delta \right)
\left(  x-y -i\delta \right) } \right] .
\label{det2}
\end{eqnarray}
This representation is general. In order to rewrite the right-hand side of this equation in our
case one should take into account that only the region  $y \to y' \to x$ is essential
in the first term. However at the point $y \to y'$  the argument of the exponential in the
 Green function (\ref{s6a}) vanishes. This means that the contribution is
determined by the preexponential pole and only the first and the second terms of
the expansion of exponential
can give nonvanishing contributions. All singularities
under the integral over $y$ in the function coming from  the first term  are in one
semiplane. One can close the contour in the other one and prove that this integral
vanishes. In the next order in $\Phi$ only the part with the singularity in the lower semiplane
of $y $ gives a nonvanishing term. After  integration over $y'$ one has (in
momentum space representation ):
$$
{-i\over 2\pi}\int^T_0 dt_1\int^{\infty}_0 {dp\over 2\pi}p \Phi_{-p}\left( T\right)
\Phi_{p}\left( t_1\right)
\exp{ \left( -ipv_f  \left( T-t_1 \right)\right)} .
$$
The $L$-electrons give the same result but with the opposite sign of $p$  in the
region $p< 0$. After the integration of \eq{det2} and symmetrization one gets
\eq{s9}. Note that  \eq{s9} is  gauge invariant: the fields depending only on
 time  do not contribute  to \eq{s9}.

\section{ Normalization coefficient and  energy shift.}
\label{z}

We have seen that the matrix element from eq.(\ref{s1}) can be expressed as
a Gaussian type functional integral. It gives the normalization coefficient and the
ground state energy shift. Indeed, one can expand the exact wave function over
the free electrons ones. In the limit $T \to \infty$ only the matrix element between
the lowest energy level survive. It can be represented in the form:
$$
Z= \exp{\left( -i\Delta E T\right) }|<\Omega|F>|^2
$$
where $\Delta E$ is the ground state energy shift. Comparing $Z$
with the definition of the normalization coefficient
$Z_0$ from
eq.(\ref{rt5}) one can see that it is equal to the overlap
probability of the ground states of  the free and interacting
electrons~: $|<\Omega|F>|^2.$ The normalization coefficient should
be calculated for a finite size system, as it is exponentially
small with the volume.

On the other hand, the matrix element
we are interested in is
\begin{eqnarray}
Z=1/ {\cal N}\int {\cal D}\Phi \exp \left[ {i\over2} \int_0^T dt dt_1
\int^\infty _{-\infty}{dp\over2\pi}
\Phi\left( p,t \right) \Phi\left( -p,t_1 \right) V^{-1} \left( p  \right)
\delta \left( t-t_1  \right) \right.
\nonumber\\
- {1\over 4\pi } \int _0^T dt dt_1 \int _{-\infty}^\infty {dp\over 2\pi}
 \Phi \left( -p,t  \right) \Phi \left( p,t_1  \right) |p| \exp\left[
 -i|p|v_f |t-t_1|  \right] ,
\label{z1}
\end{eqnarray}
where $1/{\cal N}$ is the normalization coefficient (\ref{s10a}).

It is convenient to transform integral operator  The  more effective procedure is the transformation of the
integral operator  (\ref{z1}) to a differential form. In order to do this we note the
identity
\begin{equation}
{1\over -2i|p|v_f}\left( {\partial^2 \over \partial t^2}+p^2v_f^2  \right)
\int_0^T dt_1\Phi _p\left( t_1  \right)
\exp{\left( -i|p| v_f|t-t_1| \right)}= \Phi_p\left( t \right) .
\label{z2}
\end{equation}
Thus symbolically
$$
\exp{\left( -i|p| v_f|t-t_1| \right)}={-2i|p|v_f \over \left( {\partial^2 \over
\partial t^2}+p^2v_f^2  \right)}
\delta \left( t-t_1 \right) ,
$$
and the kernel of \eq{z1} is equal to:
\begin{equation}
{i \over 2V \left( p  \right)}  {{\partial^2 \over \partial t^2} +\omega_p^2\over
{\partial^2 \over \partial t^2} +p^2v_f^2} \delta \left( t-t_1  \right).
\label{z3}
\end{equation}
As a result, one has:
\begin{eqnarray}
Z=1/ {\cal N}\int {\cal D}\Phi \exp \left[ {i\over2} \int_0^T dt
\int^\infty _{-\infty}{dp\over2\pi}V^{-1} \left( p  \right)
\Phi\left( -p,t \right){{\partial^2 \over \partial t^2} +\omega_p^2\over
{\partial^2 \over \partial t^2} +p^2v_f^2} \Phi\left( p,t \right)   \right]  .
\label{z4}
\end{eqnarray}
Taking into account that the normalization coefficient ${\cal N}$ cancels the
$\sqrt{Det \left( i/2V_p  \right)}$
that arises from the  differential kernel definition one gets:
$$
Z^{-2}=Det \left[  {i\left( { \partial^2 \over \partial t^2} +\omega_p^2 \right) \over
i\left( {\partial^2 \over \partial t^2 } +p^2v_f^2 \right) } \right] = {D\over D_0} .
$$
In order to define the differential operator one should have two initial conditions.
In exactly the same way as for  the derivation of  the equation for the saddle
point field one gets this conditions:
\begin{eqnarray}
\partial_t \Phi \left( p, 0  \right) -i|p|v_f\Phi \left( p, 0  \right) = 0
 \nonumber \\
\partial_t \Phi \left( p, T  \right) +i|p|v_f\Phi \left( p, T  \right) = 0 .
\label{z4a}
\end{eqnarray}

Usually, one calculates determinants  with zero boundary conditions
$$\Phi_p\left( 0 \right)= \Phi_p\left( T \right)=0 $$
In order to reduce our problem to the problem with zero boundary conditions
let us introduce
$$ \Phi_p\left( t  \right) = {\overline \Phi_p\left( t  \right) }+ \phi\left( t  \right). $$
The field ${\overline \Phi_p\left( t  \right) }$ is supposed to obey the equation
$\bigtriangleup_t {\overline \Phi_p\left( t  \right) }= 0$, with   initial
conditions \eq{z4a}. (As usual $\bigtriangleup_t ={ \partial^2 \over \partial t^2} +\omega_p^2$.)
The field $\phi\left( t  \right)$ is arbitrary but with zero boundary conditions.
The solution is
${\overline \Phi_p\left( t  \right) }$ can be expressed in the following form:
$$
{\overline \Phi_p\left( t  \right) } =  \Phi_p\left( 0 \right) {\sin{\omega_p\left(T- t
\right)}\over
\sin{\omega_pT}}
+  \Phi_p\left( T  \right){\sin{\omega_p\left( t  \right)}\over
\sin{\omega_pT}}.
$$  (Constants $\Phi_p\left( 0 \right)$ and $\Phi_p\left( T \right)$ are arbitrary.)
This means that the determinant has the form
\begin{eqnarray}
D^{-1/2}= \int^{\infty}_{-\infty} d\Phi_p\left( 0 \right) d\Phi_p\left( T\right)
\exp{\left[ i\left( \Phi_p\left( T \right)
\partial_t\phi_p\left( T \right) - \Phi_p\left( 0 \right)\partial_t\phi_p\left( 0
\right)\right)\right]}
\nonumber \\
\int {\cal D}\phi_p\left( t \right) \exp{\left( \phi_{-p}\left( t \right)\bigtriangleup_t
\phi_p\left( t \right)\right)} .
\label{z5}
\end{eqnarray}
The integral over $\phi_p$  can be calculated in the usual way:
$$
C\left( p=0 \right) {\omega_p\over \sin{\omega_pT  }},
$$
where $C\left( p=0 \right) $ is the $p=0$ contribution. It will cancel in the final
expression. Taking into account the identity
\begin{eqnarray}
\Phi_p\left( T \right)
\partial_t\phi_p\left( T \right) - \Phi_p\left( 0 \right)\partial_t\phi_p\left( 0 \right) =
 -\left(
\Phi_p\left( 0 \right)^2 +\Phi_p\left( T \right)^2\right) \left( |p|v_f  - i\omega_p
\cot{\omega_pT}
\right)+ \nonumber \\{2i\omega_p\over \sin{\omega_pT}}
\Phi_p\left( 0 \right)\Phi_p\left( T \right) , \nonumber
\end{eqnarray}
one has
\begin{eqnarray}
Z^{-2}= \prod_{p\ne 0} {\sin{\omega_pT}\over \sin{|p|T}} {p^2v_f^2  - 2i\omega_p|p|v_f
\cot{\omega_pT}
+\omega_p^2\over 2|p|v_f\omega_p\left( 1- i\cot{\omega_pT}\right) }.
\label{z6}
\end{eqnarray}
If temperature is non-zero one should substitute $T$ by $1/\Theta$. Note that \eq{z6}
is valid even at the temperature $\Theta\ll \Theta_{chiral},$ because only the Green functions
with equal time arguments were used. In this temperature region
the $Z$ can be expressed in the following form:
\begin{equation}
Z=\prod_{p\ne 0}\exp{\left( -{\omega_p-|p|v_f\over 2\Theta}\right)}{4\sqrt{|p|v_f\omega_p}\over
\omega_p+|p|v_f}
\label{z7}
\end{equation}
It is convenient to rewrite this equation in the following way:
\begin{equation}
Z= \exp{\left[ -{L\over \Theta}\int^{\infty}_{0}{dp\over
2\pi}\left( \omega_p-pv_f \right) +1/2 \sum_{p\ne
0}\log{{4\sqrt{|p|v_f\omega_p}\over \omega_p+|p|v_f} }\right]} .
\label{z8}
\end{equation}
In this form one can see the energy shift (first term in the
exponent)and  normalization coefficient (second term)
explicitly. The sums in this equation diverge because of the
gapless spectrum. They have to cut off at $p_{max}\sim 1/d$. In
order to take the preexponential factor one should calculate the
next correction  after the Riemann sum. As a result we have for the short
range potential
$$
\Delta E \sim {L \over 4\pi d }{v_f\over d} \sqrt{V_0/\pi v_f}
$$
for the energy shift and
$$
Z_0 = 4 \sqrt{\pi v_f/V_0}\exp{\left( -{L\over 4\pi d}
\log{V_0/\pi v_f}\right) }
$$
for the normalization coefficient.

\section{Calculation of Sums in Section \ref{vo}}
\label{sum}

All the sums in the equation for the action can be calculated by differentiation
of the expression
$S\left( \alpha \right)$ with respect to parameter $\alpha$:
$$
S\left( \alpha \right) = - {2\pi\over L} \sum_{n_{min}}^{\infty} {1\over p_n}
\exp{\left[ {2\pi in\alpha\over L}\left( x  +i\delta\right) \right]} ,
$$
( $\alpha$ varies within the region $\left(1, i\infty\right)$).  After summation
of the geometric series one can rewrite it in the following form:
$$
S\left( 1, x \right) = \int^{1-y_0}_1 {dy\over y} \left( 1 -y\right) ^{n_{min} -1} ,
$$
where $y_0 \left( x \right) = \exp{\left( 2\pi i/ L \right)}\left( x +i\delta \right) .$
It is implied that $x\ll L$ here.  This result can  apply at $x\sim L$ as an order of
magnitude estimate only.
The final expression appearing in the action is:
\begin{equation}
S\left( 1, x \right) - S\left( 1,y \right) = \int^{\left(  x+i\delta \right)}_{\left(
y+i\delta \right)}
{dz\over z}\exp{\left(-{iz\over\zeta }\right)} ,
\label{sum1}
\end{equation}
where $\zeta =L/ 2\pi \left( n_{min}-1 \right)^{-1}.$ If $ n_{min}\sim L\Theta/2\pi v_f^c$ then
$\zeta$ is
equal to the coherence length \eq{t8a1}.

Let us consider the influence of the boundary conditions  on  the action. In principle,
any of them can be rewritten as $p_n={2\pi}\left( n +\delta n \right) /L,  |\delta n| <1/2.$
In this case at $\Theta=0$ the action is determined by the sum:
$$
S'\left( \alpha \right) = - {2\pi\over L} \sum_{1}^{\infty} {1\over p_n}
\exp{\left[ {2\pi i\left( n+\delta n\right)\alpha\over L}\left( x  +i\delta\right) \right]} .
$$
In the same way one gets
$$
S'\left( 1, x \right) - S'\left( 1,y \right) = \int^{\left(
x+i\delta \right)}_{\left(  y+i\delta \right)} {dz\over
z}\exp{\left( -{2\pi iz\delta n\over\ L }\right)} .
$$
The result of this is that up to $|x-y |\sim L$ at $\Theta=0$ the action does not depend on the
boundary conditions. For ($\Theta \gg \Theta_c$) one should cut off
the sum at some $n=n_{min}.$  As a result $\zeta$ is substituted by
$L/ 2\pi \left( n_{min} +\delta n-1 \right)^{-1}$ in  \eq{sum1}. This suggests the
 replacement of $\Theta_c$ by $\left( 1-\delta n \right) \Theta_c $  \cite{TC}. However
the transition temperature can be defined only up to a factor of the order of unity.
Therefore we should not take this into account.


\begin{thebibliography}{99}
\bibitem{par}
In order to draw an analogy between exiton-like pair and the bound state one should
take into  account that we have a correlation between the filled states rather than a real
bound pair.
\bibitem{Back}
Let us note that this case includes also backscattering of
electrons with transition $L\to R$ and back:
\[
\int dx \hat \Psi^{\dag}_R(x) \Psi_L(x) \hat \Psi^{\dag}_L(x)\Psi_R(x).
\]
By anticommuting the $\hat \Psi$-operators one can reduce this term to the
term without backscattering \cite{E}, i.e. for this model the
Tomonaga-Luttinger Hamiltonian describes the entire  electron-electron
interaction.
\bibitem{Bohr}
In order to obtain the real parameter one should use usual expression
for the radius of Debye screening
$ R_D= 1/\sqrt{4\pi e^2{ \partial n\over\partial \mu}}$ with concentration
$n={p_f/ \pi^2d^2}.$ We took into account here that
there is only one state for electrons in directions perpendicular
to the channel. Thus $R_D\sim {d\over 2}\sqrt{\pi p_f a_b}$  where
$a_b=\left( me^2 \right) ^{-1} $ is the Bohr radius. Hence. condition
$R_D\ll d$ is equivalent to $p_fa_b\ll 1$. The latter
parameter depends only on the effective mass of the electron. If it is
of the order of the  mass of the free electron then
$a_b\sim 0.5\cdot  10^{-8}cm$  and for the concentrations typical for
metals we get $p_fa_b\ll 1$. In other words, in this case we deal
with a short range interaction with potential $V=V_0\delta(x-y)$.
\bibitem{cate}
Here we take into account that the  exact Hamiltonian
by itself contains a natural cutoff. Indeed let us
consider terms with transition of R electrons to L ones and back:
$$\begin{array}{rcl}
 \int _{-\infty}^{\infty} dx dy \exp \left( -2ip_f \left( x-y)\right)
\right) \hat \Psi^{\dag}_R
\left( x \right) \hat \Psi _L\left( x  \right) V\left( x-y \right)
\hat \Psi^{\dag}_L\left( y \right) \hat \Psi _R\left( y  \right) .\\
 \end{array}
$$
They correspond to large momentum transfer: the argument of
potential $2p_f-p$ is close to Fermi momentum. Neglecting $p$ as
compared to $p_f$ and proceeding back to the coordinate representation
we see that these terms look like
$
-V\left( 2p_f \right)\delta \left( x-y \right)
$ .
Now it is easy to check that if $V(p)$
is the Coulomb potential with some cutoff then the original
cutoff is canceled out
in the sum of all terms and substituted by the value of the $2p_f$.
\bibitem{matrix}
Strictly speaking at non-zero temperatures the system is described not
by the wave function but by the density matrix. However we are interested in
the low temperatures region where the probability to find  system in the
excited state  is  small. In the main approximation one can describe such a system by
a wave function.
\bibitem{rem7}
In order to calculate this one should take into account the identity
\begin{equation}
\left[ \hat a \left( p \right) , \exp{ \left( \int {dp' \over 2\pi
}\hat a^{\dag} \left( p' \right) \hat b^{\dag}  \left( p' \right)
\right) } \right]  =
\hat b^{\dag} \left( p \right) \exp{ \left( \int {dp' \over 2\pi
}\hat a^{\dag} \left( p' \right) \hat b^{\dag}  \left( p' \right)
\right)} \nonumber
\end{equation}
and determine the anticommutator with  same momentuma in a  regular way:
$ \{ \hat a^{\dag} \left( p_n \right) ,\hat a \left( p_m
\right) \} = L \delta_{n,m} .$
\bibitem{rem4}
The bosonization technique allows one to calculate the 4-particle correlator exactly ( i.e.
with pairs scattering). As a result, one has the well-known power:
$$ \alpha_T=1 -1/\left[ 1+\sqrt {1 +{V_0\over \pi v_f}} \right]   .$$
It coincides with $\alpha_0 $  only in the  strong interaction
limit. This distinction arises from the fact that in the term we
are interested in only the direct interaction between two pairs
enters.
\bibitem{rem2a}
The terms in the equation for ${\cal S}_{eff}$ having the
factor
$\exp{-|p_n|v_f^c\over \Theta}$ describe an excited state.
We do not discuss them  because one can calculate such effects
more easily by using the bosonization technique.
\bibitem{rem5}
Such behaviour of the density-density correlators is well-known
(see e.g.\cite{Shul}.)  Note
that backscattering processes not included into the Hamiltonian in
this case modify the numerical coefficient  only in the $\log $.
\bibitem{TC}
This means that $\Theta_c$ is determined by the excitation energy
with the smallest momentum.
\end{thebibliography}
\end{document}